\documentclass[conference]{IEEEtran}
\IEEEoverridecommandlockouts
\usepackage{tikz}
\usepackage{hyperref}
\usepackage{color,soul}

\usepackage{graphicx}
\usepackage{textcomp}
\usepackage{xcolor}
\usepackage{xspace}
\usepackage{lipsum}
\usepackage{multirow}
\usepackage{booktabs}
\usepackage[noend]{algorithmic}
\usepackage[ruled,vlined,linesnumbered,algo2e]{algorithm2e}
\usepackage{makecell} 

\usepackage{cite}
\usepackage{amsmath,amssymb,amsfonts}
\usepackage{graphicx}

\newcommand{\eva}{\textit{EVA}\xspace}

\newcommand{\Fig}[1]{Fig.~\ref{#1}}

\newcommand{\Tbl}[1]{Tbl.~\ref{#1}}
\newcommand{\Sec}[1]{Sec.~\ref{#1}}
\newcommand{\Reb}[1]{\textcolor{red}}

\definecolor{mycolorblue}{RGB}{81,140,230}
\definecolor{mycolorgreen}{RGB}{166,213,95}
\definecolor{mycoloryellow}{RGB}{242,210,98}
\definecolor{mycolorpink}{RGB}{242,96,119}
\definecolor{mycolorpurple}{RGB}{149,90,189}
\definecolor{mycolororange}{RGB}{255,199,7}

\newcommand\circlednumberblue[1]{%
  \begin{tikzpicture}[baseline=(char.base)]
    \node[shape=circle,,fill=mycolorblue,inner sep=1pt] (char) {\textcolor{white}{\scriptsize\sffamily\bfseries#1}};
  \end{tikzpicture}}

\newcommand{\revise}[1]{#1}

\newcommand{\resp}[2]{#2}

\def\BibTeX{{\rm B\kern-.05em{\sc i\kern-.025em b}\kern-.08em
    T\kern-.1667em\lower.7ex\hbox{E}\kern-.125emX}}
\begin{document}

\pdfpagewidth=8.5in
\pdfpageheight=11in

\newcommand{\iscasubmissionnumber}{75}

\pagenumbering{arabic}


\title{EVA: Accelerating LLM Decoding via an Efficient Vector Quantization Architecture}
\author{
    \IEEEauthorblockN{
        Bowen Duan\textsuperscript{\dag*}, 
        Cong Guo\textsuperscript{\dag*\S}, 
        Chiyue Wei\textsuperscript{\dag}, 
        Haoxuan Shan\textsuperscript{\dag}, 
        Yuzhe Fu\textsuperscript{\dag}, 
        Xinhua Chen\textsuperscript{\dag}, \\
        Yifan Xu\textsuperscript{\dag}, 
        Ziyue Zhang\textsuperscript{\dag}, 
        Changchun Zhou\textsuperscript{\dag}, 
        Hai Li\textsuperscript{\dag}, 
        Yiran Chen\textsuperscript{\dag}
    }
    
    \vspace{1ex}
    
    \IEEEauthorblockA{
        \textsuperscript{\dag}Duke University, USA \quad 
        \textsuperscript{*}Equal contribution \quad 
        \textsuperscript{\S}Corresponding author \\
        \{bowen.duan, cong.guo, hai.li, yiran.chen\}@duke.edu
        \vspace{-12pt}
    }
}

\maketitle
\thispagestyle{empty}
\pagestyle{empty}

\begin{abstract}
Large Language Models (LLMs) have achieved impressive performance across diverse domains but remain inefficient during the autoregressive decoding phase. Unlike the prefill stage, which employs compute-bound GEMM operations, decoding executes a sequence of small GEMV-like computations that are memory-bound and underutilize modern accelerators. Weight-only vector quantization (VQ) has emerged as an effective compression technique that clusters model weights into a shared \textit{codebook} and replaces the original weight matrix with low-precision \textit{indices}, enabling 2-bit-level weight compression. While this approach substantially reduces model size and memory bandwidth, it still suffers from two critical inefficiencies: the low utilization of GEMV computation and frequent memory conflicts during codebook lookups. 

This paper presents \textbf{\eva{}}, an efficient vector-quantization-based architecture that addresses both computational and memory bottlenecks in LLM decoding. \eva{} builds on a simple yet effective insight that combines input-codebook computation with conflict-free memory access. Instead of reconstructing quantized weights from indices, \eva{} directly performs dot products between input vectors and the weight codebook, transforming LLM decoding from GEMV to GEMM computation. It then performs structured lookups from an intermediate output buffer, eliminating memory bank conflicts. We further design a hardware-software co-optimized architecture specialized for LLM decoding while remaining compatible with conventional prefill execution. Evaluations show that \eva{} achieves up to \textbf{11.17$\times$} speedup and \textbf{7.17$\times$} higher energy efficiency compared with the SOTA lookup-based architecture, while preserving arithmetic precision after vector quantization. Our code is available at \url{https://github.com/dbw6/Eva.git}.
\end{abstract}

\section{Introduction}

\begin{figure}[t]
    \centering
    \includegraphics[width=0.48\textwidth]{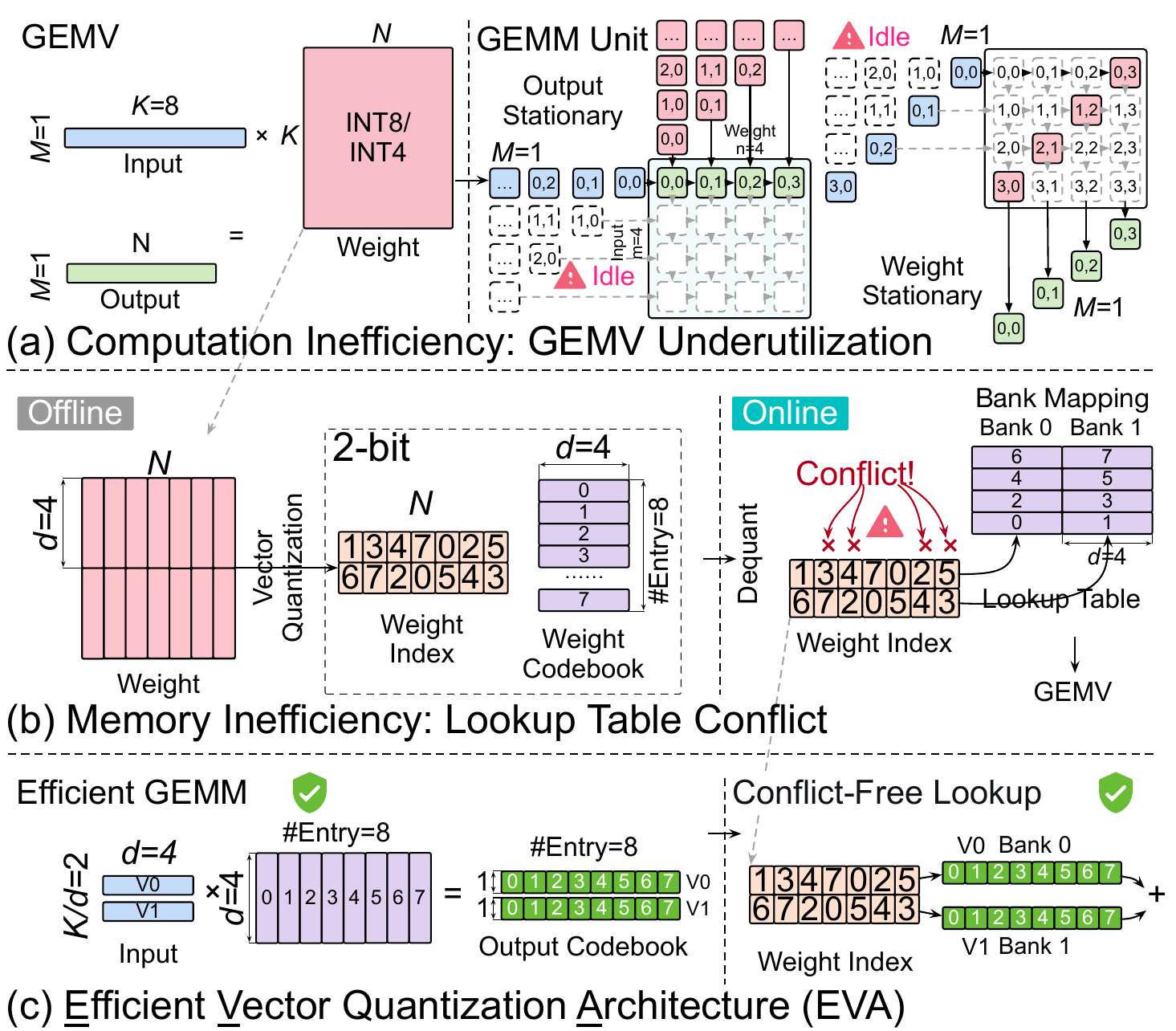}
    \caption{
    Motivation of this work.  
    (a) \resp{RD-Q2}{Conventional GEMV suffers from poor utilization of compute units during LLM decoding.} 
    (b) Vector quantization introduces irregular and conflicting memory accesses during online lookup.  
    (c) \eva{} eliminates lookup conflicts and reformulates LLM decoding into GEMM-style computation, enabling efficient execution on modern accelerators.
    }
    \label{fig:intro}
    \vspace{-10pt}
\end{figure}

Large Language Models (LLMs)~\cite{Devlin2018BERT,touvron2023llama2,radford2019language_gpt2, bai2023qwen,lin2024speechprune} have achieved remarkable success across a wide range of domains, from natural language understanding~\cite{deepseek2025deepseek_r1,OpenAI2023GPT4} to code generation~\cite{thakur2024verigen,github_copilot, wang2025angles}.  
However, the inference efficiency of LLMs remains a critical challenge, particularly in the \textit{decoding phase} of the autoregressive (AR) mechanism~\cite{vaswani2017attention}.  
During inference, computation is divided into two distinct phases: \textit{prefill} and \textit{decoding}.  
The prefill phase processes the entire input sequence and can be efficiently executed as large-scale General Matrix Multiplication (GEMM) operations.  
In contrast, the decoding phase generates one token at a time, resulting in a sequence of small-scale General Matrix--Vector Multiplication\revise{-like} (\revise{GEMV-like}) operations.  
As illustrated in \Fig{fig:intro} (a), this fundamental difference in computational granularity makes the decoding phase significantly less efficient.

Quantization has become an essential technique for efficient LLM inference~\cite{jacob2018quantization,guo2022ant,guo2023olive,hu2025m,lin2024qserve, lin2024awq, navardi2025genai}.  
Conventional quantization methods~\cite{jacob2018quantization,hu2024llm,kim2021bert} typically apply low-precision formats to both weights and activations, enabling reduced-precision computation and lower memory access cost.  
Recent frameworks primarily focus on \textit{weight-only quantization}~\cite{lee2025amq, wang2025dobi, frantar2022gptq, wang2023bitnet}, such as AWQ~\cite{lin2024awq} and SqueezeLLM~\cite{kim2024squeezellm}.  
These methods compress model weights into low-bit representations (e.g., 4-bit) while retaining high-precision computation (e.g., FP16) for activations.  
Weight-only quantization directly targets the \textit{decoding} bottleneck of LLMs, where activations are lightweight but weight tensors dominate memory traffic.  
Although activations remain in high precision to preserve model accuracy, weight-only quantization achieves nearly linear speedup in the decoding stage as the weight precision decreases~\cite{lin2024awq}.
This improvement arises because LLM decoding is inherently memory-bound, and quantization effectively mitigates the bandwidth bottleneck caused by frequent weight loading.

Pushing this trend further leads to \textit{codebook-based compression}~\cite{cheng2025ecco,han2015deep,egiazarian2024extreme}, where weight tensors are represented using lookup tables (LUTs) instead of arithmetic quantization functions.  
Unlike integer or floating-point quantization, lookup-based quantization provides higher flexibility and better fidelity by enabling non-uniform representations.  
Recently, \textbf{\textit{Vector Quantization} (VQ)}~\cite{tseng2024quip,egiazarian2024extreme,huijben2024residual,gou2025carvq, liu2024vptq} has emerged as a pivotal technique that extends codebooks from single-element (scalar) to multi-element (vector) representations, allowing each codebook entry to encode multiple weight elements (e.g., 4 or 8).  
Compared with arithmetic quantization methods such as AWQ~\cite{lin2024awq}, which are typically limited to 4-bit precision, VQ pushes this limit further to 2-bit quantization while maintaining high accuracy, demonstrating superior algorithmic efficiency.  
As illustrated in \Fig{fig:intro} (b), an example with a vector size of $d=4$ partitions the weight matrix into 4-element vectors, which are then mapped to a compact weight index (WI) matrix and a weight codebook (WC).  
This multi-dimensional compression enhances quantization expressiveness and achieves state-of-the-art accuracy-compression trade-offs while significantly reducing model size.

Despite its effectiveness in model compression, conventional VQ offers \textbf{no speedup} on current hardware accelerators (e.g., GPUs); in fact, it can even be slower than FP16 inference~\cite{liu2025vq}.  
This limitation arises from two fundamental challenges:

\textbf{(1) Compute inefficiency (\Fig{fig:intro} (a)).}  
While VQ effectively reduces memory bandwidth requirements, it does not alter the computational structure of the decoding phase, which remains a sequence of small-scale GEMV operations.  
Such matrix--vector computation is inherently inefficient on modern accelerators optimized for large GEMM workloads, leading to low parallelism and poor utilization. This inefficiency is not unique to VQ but is a general limitation of weight-only quantization in LLM decoding~\cite{agrawal2023sarathi,patel2024splitwise}.

\textbf{(2) Memory inefficiency (\Fig{fig:intro} (b)).}  
Even after compressed weights are loaded into on-chip memory, the ensuing \textit{dequantization} via codebook lookup often suffers from severe memory access conflicts.  
These conflicts stem from the irregular and uncoalesced indexing patterns inherent to LUT-based quantization, which lead to serialization, limited memory throughput, and reduced cache efficiency.  
For example, as shown on the right side of \Fig{fig:intro} (b), indices 5 and 3 attempt to access the same memory bank simultaneously, resulting in a memory conflict and serialized access.  
This bottleneck fundamentally limits their attainable speedup, regardless of their impressive compression ratios.

To overcome these challenges, we propose \textbf{\eva{}}, an \underline{E}fficient \underline{V}Q-based \underline{A}rchitecture that addresses both computational and memory inefficiencies in the LLM decoding phase.  
\eva{} is built on a simple yet effective \textit{aha insight}, consisting of two main steps that jointly restructure computation and remove memory conflicts.

\textbf{Step 1: From weight decoding to codebook dot product.}  
Since each vector in the VQ-quantized weight matrix originates from the codebook, explicit reconstruction is unnecessary.  
As shown on the left side of \Fig{fig:intro} (c), instead of decoding the weight indices (WIs) on-chip to reconstruct the full weight matrix, we directly compute dot products between the input vectors and the weight codebook (WC), generating intermediate results that collectively form an \textit{output codebook} (OC).  
To support this operation, the input is partitioned into multiple vectors (e.g., $v_0$ and $v_1$), each multiplied by the WC to produce corresponding intermediate results that together constitute the OC.

\textbf{Step 2: Conflict-free lookup from the output codebook.}  
As shown on the right side of \Fig{fig:intro} (c), the final outputs are obtained by performing lookup operations on the OC using the WI matrix.  
Unlike conventional VQ, where lookups in the WC frequently cause memory bank conflicts, this formulation eliminates such conflicts entirely.  
During OC computation, matrix multiplication implicitly distributes OC elements across different banks, as each bank stores results derived from distinct input vectors.  
For example, when indices such as 5 and 3 in the rightmost column of the WI matrix are accessed simultaneously, they map to different banks (corresponding to $v_0$ and $v_1$), enabling fully parallel and conflict-free access.

\textbf{Advantages.}  
This approach improves both computational and memory efficiency.  
(1) \textit{Computation.}  
The LLM decoding process is effectively \textbf{reformulated from GEMV to GEMM}, increasing arithmetic intensity and parallelism.  
Since the codebook is small (e.g., 256 entries) while $N$ is large (e.g., 4096), the overall computation cost is significantly reduced.  
(2) \textit{Memory.}  
Compared with conventional VQ, accessing the same number of weight indices does not require additional memory banks.  
At the same time, the bandwidth requirement is reduced from reading $d=4$ weight elements per access to reading a single OC element, and memory conflicts are completely eliminated.  
These optimizations make \eva{} efficient in both computation and memory utilization.

In summary, \eva{} transforms the inefficient VQ decoding process into a conflict-free GEMM-style computation.  
We further design a dedicated hardware architecture that implements this reformulation to achieve high utilization and scalability based on a simple yet effective insight.

Specifically, {\eva} makes the following contributions:

\begin{itemize}
    \item We propose a \textbf{codebook-driven GEMM formulation} that replaces conventional weight decoding with direct dot products between input vectors and the weight codebook.  
    This formulation transforms memory-bound GEMV operations into compute-efficient GEMM computations, significantly improving utilization.

    \item We develop an \textbf{output-codebook lookup mechanism} that reorganizes memory access into a conflict-free structure.  
    By distributing output codebook entries across memory banks, \eva{} eliminates lookup conflicts and reduces bandwidth demand.

    \item We design and implement a \textbf{hardware--software co-optimized architecture} tailored for efficient LLM decoding while maintaining compatibility with conventional accelerator execution in the prefill stage.
\end{itemize}

Extensive experiments demonstrate that \eva{} achieves up to \textbf{11.17$\times$} speedup and \textbf{7.17$\times$} improvement in energy efficiency over the state-of-the-art lookup-based baseline, FIGLUT, while preserving arithmetic precision after vector quantization.
\section{Background and Motivation}

\subsection{LLM Decoding: Computational Inefficiency}
The core operation in transformer-based Large Language Models (LLMs)~\cite{touvron2023llama2,jiang2023mistral_7b} is matrix multiplication, expressed as $\mathbf{Y} = \mathbf{X}\mathbf{W}$.  
Here, $\mathbf{X} \in \mathbb{R}^{M \times K}$ denotes the input matrix, and $\mathbf{W} \in \mathbb{R}^{K \times N}$ represents the projection or feed-forward weight matrix.  
Autoregressive LLMs~\cite{vaswani2017attention} are typically executed in two distinct phases: \emph{prefill} and \emph{decoding}.  
During the prefill phase, the model processes multiple tokens from several sequences simultaneously~\cite{yu2022orca}.  
As a result, all three dimensions ($M$, $K$, and $N$) of the matrix multiplication are large, enabling General Matrix Multiplication (GEMM).  
This configuration allows GPUs and other accelerators to exploit extensive data reuse, thereby achieving high computational efficiency~\cite{choquette2021nvidia_a100,patel2024splitwise, jouppi2017datacenter}.

In contrast, during the \emph{decoding} phase, the model generates tokens sequentially, one at a time.  
Consequently, the total number of rows is reduced to $M = 1$, turning the matrix multiplication $\mathbf{X}\mathbf{W}$ from a large-scale GEMM operation into a small-scale General Matrix--Vector Multiplication (GEMV) operation.  
As shown in \Fig{fig:intro} (a), performing GEMV on accelerators optimized for GEMM is highly inefficient for two primary reasons.  
First, when $M$ is small, only a narrow portion of the GEMM unit is active, while most processing element (PE) lanes remain idle.  

Second, GEMV inherently suffers from low arithmetic intensity because one dimension of the matrix cannot be reused across computations.  
During decoding, activations are small, but the weight matrix dominates the memory footprint and must be repeatedly fetched for each generated token without reuse.  
As a result, GEMV operations become memory-bound, with the majority of runtime dominated by weight loading rather than computation.  
This inefficiency has motivated extensive research on compressing the weight matrices during the decoding process to reduce memory bandwidth demand and improve overall inference efficiency~\cite{lin2024awq,lin2024qserve}. 

\resp{RC-C3}{GEMVs occur only in the decode phase with a batch size of 1. In this article, we will refer to both strict GEMVs and GEMV-like GEMMs from small batches, as they both cause array underutilization.}

\begin{figure}[t]
    \centering
    \includegraphics[width=0.47\textwidth]{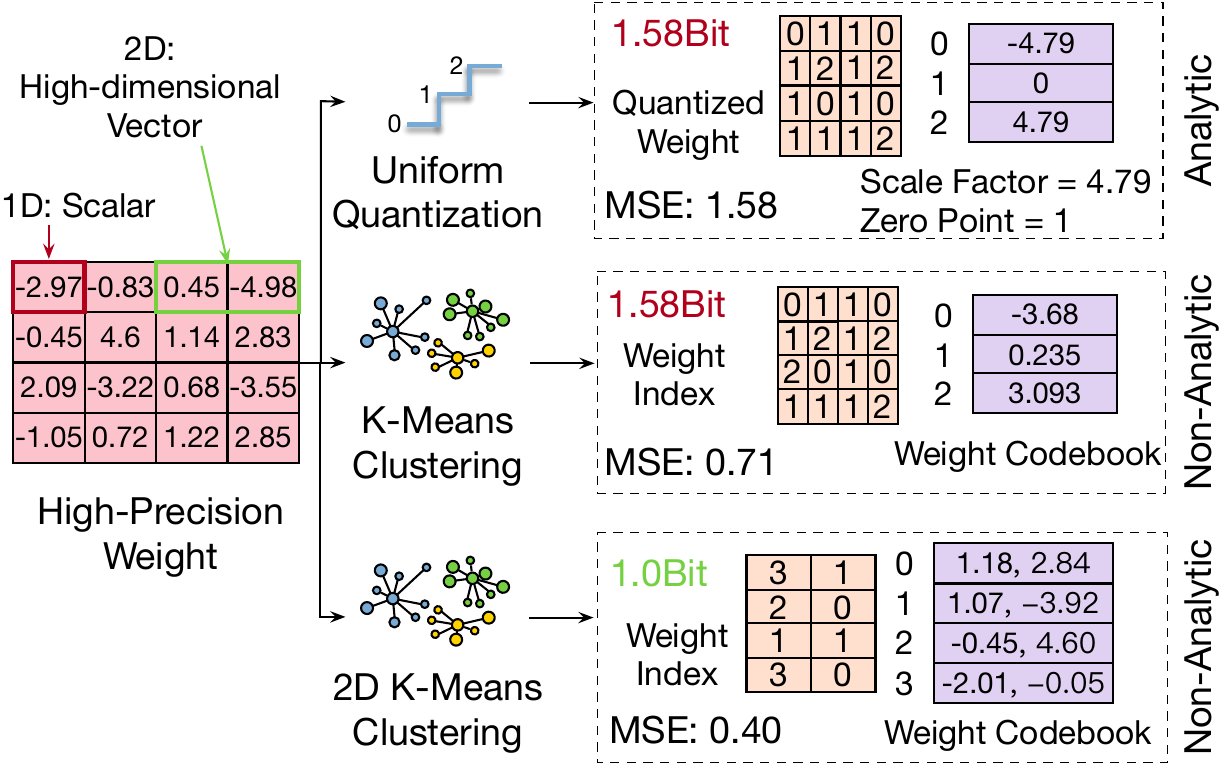}
    \caption{
    Comparison of quantization schemes:
    (a) uniform quantization (analytic); 
    (b) K-means (1D vector) quantization (non-analytic); 
    (c) 2D vector quantization (non-analytic).
    }
    \label{fig:bg}
    \vspace{-10pt}
\end{figure}

\subsection{Quantization}

Quantization methods for LLMs can be broadly categorized into two classes:
\textbf{\textit{analytic quantization}} and \textbf{\textit{non-analytic quantization}}.  
As illustrated in \Fig{fig:bg}, the key distinction lies in whether the quantization function can be explicitly represented by arithmetic operations (e.g., rounding, scaling).  

\textbf{Analytic quantization.}  
As shown in \Fig{fig:bg} (a), analytic quantization refers to schemes with explicit mathematical formulations that map high-precision values (e.g., FP32, FP16) to low-bit values (e.g., INT4/8, FP4/8) through simple operations such as linear scaling and rounding.  
Both activations and weights can be quantized in this way, and such approaches are widely adopted for compute-bound workloads where reduced-precision arithmetic directly lowers FLOPs~\cite{yao2022zeroquant,kim2021bert}.  

However, the decoding phase of LLM inference is inherently \textit{memory-bound}, rather than compute-bound.  
Therefore, modern systems typically employ \textbf{weight-only quantization}~\cite{lin2024awq,frantar2022gptq}, where weights are stored in low-bit representations (e.g., 4-bit), while activations remain in high precision (e.g., FP16).  
This strategy effectively reduces memory bandwidth requirements and achieves near-linear speedups proportional to the reduction in data transfer volume.  

Nonetheless, the computation itself still operates in high precision.  
Thus, this approach functions primarily as a \textit{compression technique} rather than a low-precision computation method.  
Because of its simple and analytic nature, decoding from such quantized weights is efficient: element-wise dequantization can be performed \textit{in-place} without additional dependencies.  
Consequently, weight-only analytic quantization has been widely adopted in decoding-optimized frameworks such as AWQ~\cite{lin2024awq} and GPTQ~\cite{frantar2022gptq}.  

Despite its practicality, analytic quantization offers limited representational flexibility.  
Its quantization error remains relatively high under aggressive compression.  
More importantly, \textbf{it cannot alter the underlying GEMV computation paradigm}: the decoding process remains a sequence of small, memory-bound matrix--vector multiplications.  

\textbf{Non-analytic quantization.}  
In contrast, non-analytic quantization removes the requirement for closed-form quantization functions.  
Instead, it directly learns quantization mappings by minimizing a reconstruction loss, typically the mean squared error (MSE), commonly through K-means clustering~\cite{cheng2025ecco,han2015deep,egiazarian2024extreme} (\Fig{fig:bg} (b)).  
This approach achieves substantially lower quantization errors than analytic methods and maintains high fidelity even at 2-bit precision.  
Representative schemes in this category include lookup-table (LUT)-based quantization methods~\cite{han2015deep,li2025lut,guo2022ant} that replace arithmetic dequantization with direct table lookups.  

However, this flexibility comes at a cost.  
Since the mapping between original weights and quantized representations is no longer expressible in arithmetic form, decoding requires a \textit{1-to-1 lookup} from codebooks.  
Such irregular and uncoalesced memory access patterns lead to frequent memory bank conflicts, making the decoding process both expensive and difficult to parallelize, as illustrated in \Fig{fig:intro} (b).  
This limitation has prevented non-analytic quantization from being widely adopted in large-scale inference systems.  

\textbf{Vector Quantization (VQ).}  
Vector Quantization (VQ) extends non-analytic quantization from scalar (1D) K-means to higher-dimensional clustering (multi-D K-means).  
By encoding multiple elements jointly (e.g., 4D or 8D vectors), VQ captures local structural correlations in weights and substantially reduces quantization loss.  
As shown in \Fig{fig:bg} (c), VQ achieves lower reconstruction error than analytic quantization even with an average of only 1 bit per element, reaching state-of-the-art (SOTA) accuracy-compression trade-offs.  

Nevertheless, VQ introduces substantial challenges during decoding.  
Because each vector code corresponds to multiple weight elements, the effective codebook size increases, which in turn raises memory bandwidth demands and exacerbates access irregularity.  
This “1-to-many” lookup mechanism makes VQ particularly sensitive to memory access conflicts, posing a fundamental barrier to its efficient deployment on current accelerators.  

\subsection{Codebook-based Lookup: Memory Inefficiency}
Supporting efficient codebook-based non-analytic quantization has long been a major research objective in both architecture and system communities.

\textbf{Architectural perspective.}  
From a hardware perspective, prior studies mainly focus on optimizing the lookup table (LUT) design to reduce memory conflicts.  
Representative works include GOBO~\cite{zadeh2020gobo}, FIGLUT~\cite{park2025figlut}, LUT Tensor Core~\cite{zhiwen2024luttensorcore}, and LUT-DLA~\cite{li2025lut}.  
As summarized in \Tbl{tab:lutsurvey}, these approaches typically rely on either duplicating or broadcasting the codebook across processing elements (PEs) to mitigate bank conflicts.  
However, both strategies incur substantial hardware and bandwidth overhead.  
For instance, FIGLUT broadcasts $8\times16$-bit data to a full PE column of 32 lanes, resulting in a total bandwidth requirement of $16\times 32 \times(8\times16\,\text{bits})$, while LUT-DLA requires $256\times(16\times16 \,\text{bits})$ registers to store replicated codebook entries.  
Such design costs significantly constrain scalability and limit the effective codebook size to at most 16 entries.

\begin{table}[t]
\centering
\caption{Comparison of representative LUT-based accelerator designs.}
\label{tab:lutsurvey}
\resizebox{\linewidth}{!}{
\begin{tabular}{llll}
\toprule
\textbf{Architecture} & \textbf{Codebook Size} & \textbf{Strategy} & \textbf{Replication / Bandwidth Cost} \\
\midrule
GOBO~\cite{zadeh2020gobo} & $8\times16$ bits  (FP16) & Duplication & 768 $\times$ ($8\times16$ bits) Registers\\
LUT TC~\cite{zhiwen2024luttensorcore} & $8\times 8$ bits (INT8) & Duplication & 2 $\times$ 64 $\times$ ($8\times8$ bits) Registers \\
LUT-DLA~\cite{li2025lut} & $16\times$ 16 bits (BF16) & Duplication & 256 $\times$ (16 $\times16$ bits) Registers \\
FIGLUT~\cite{park2025figlut} & $8\times 16$ bits (FP16) & Broadcast & $16\times$ 32 $\times$ $(8\times16\text{ bits})$ Bandwidth \\
\bottomrule
\end{tabular}
}
\vspace{-10pt}
\end{table}
\textbf{Software and system perspective.}  
From the algorithmic and system side, prior VQ frameworks primarily focus on improving accuracy while adopting naive software implementations.  
AQLM~\cite{egiazarian2024extreme} and QuiP~\cite{tseng2024quip} are implemented in PyTorch and mainly serve as model compression techniques, often slower than FP16 inference.  
VQ-LLM~\cite{liu2025vq} is the first GPU-based framework to address codebook conflicts; however, due to GPU hardware constraints, it only mitigates conflicts by profiling codebook access frequencies to classify \textit{hot} and \textit{cold} entries.  
This heuristic approach alleviates but does not eliminate memory contention, leaving the fundamental inefficiency of VQ unresolved.  

As a result, despite its superior compression and accuracy, vector quantization remains limited by its poor hardware efficiency and irregular access patterns.  
In this study, we propose \textbf{\eva{}}, which overcomes both computational inefficiency and memory conflicts through a unified, architecture-aware VQ design.
\section{Overview of \eva{}}
\begin{figure*}[t]
    \centering
    \includegraphics[width=0.99\textwidth]{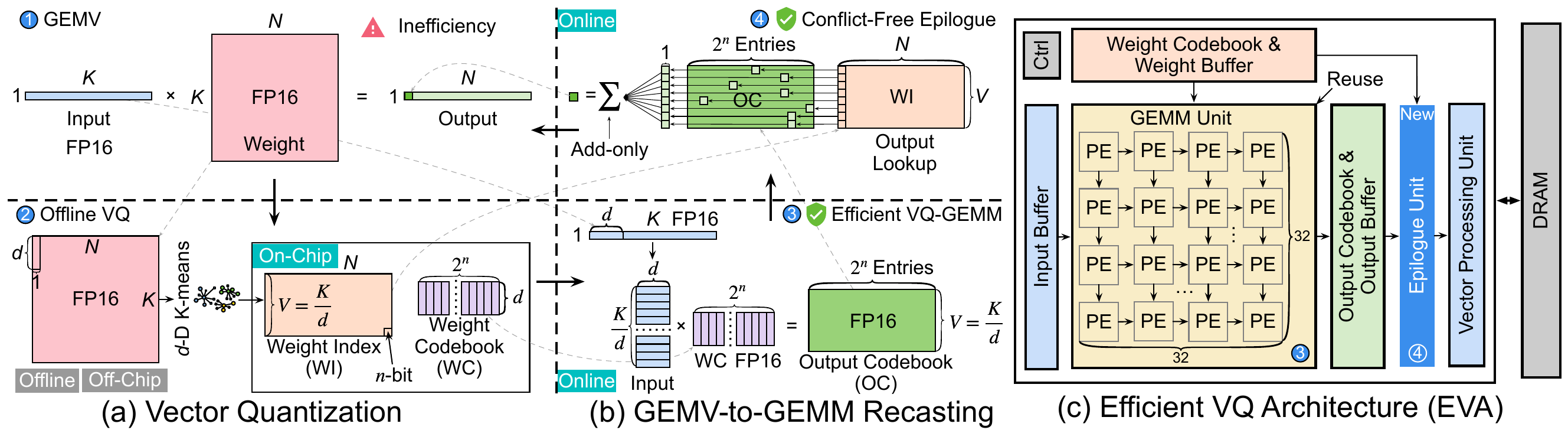}
    \caption{
    Overview of the \eva{} computation flow and architecture. 
    }
    \vspace{-10pt}
    \label{fig:main_arch}
\end{figure*}

This section presents the computation flow and architectural overview of \eva{}, 
as illustrated in \Fig{fig:main_arch}. 
We first define the key variables and symbols used throughout this study, summarized in \Tbl{tab:variables}.

\begin{table}[t]
\centering
\caption{Summary of notations.}
\label{tab:variables}
\resizebox{1\columnwidth}{!}{
\begin{tabular}{l l l}
\toprule
\textbf{Symbol} & \textbf{Description} & \textbf{Value} \\
\midrule
$M$ & Input (output) height, i.e., number of tokens & 1 \\
$N$ & Weight (output) width, i.e., output channel & 4096+ \\
$K$ & Input width or weight height for reduction & 4096+ \\
$d$ & Vector dimension used in VQ & 8 \\
$n$ & Bit-width of each vector index & 8 \\
$2^n$ & Number of entries per codebook & 256 \\
$V = \frac{K}{d}$ & Height of compressed weight index matrix & 512+ \\
$v$ & Tile height for grouped computation & 32 \\
$C$ & Number of codebooks per layer & 2--4 \\
$q = \frac{C \times n}{d}$ & Effective quantization bit-width (average) & 2--4 \\
$\mathbf{I}$ & Weight index (WI) matrix & $[0,2^n)^{V\times N}$ \\
$\mathbf{B}$ & Weight codebook (WC) containing $2^n$ centroids & $\mathbb{R}^{d\times 2^n}$ \\
$\mathbf{O}$ & Output codebook (OC) & $\mathbb{R}^{V\times 2^n}$ \\
\bottomrule
\end{tabular}
}
\vspace{-10pt}
\end{table}

\subsection{Vector Quantization}

\textbf{GEMV.}  
As illustrated in \Fig{fig:main_arch} (a)~\circlednumberblue{1},  
the LLM decoding phase performs a sequence of matrix--vector multiplications (GEMV).  
For each generated token, the computation is expressed as
\[
\mathbf{y} = \mathbf{x}\mathbf{W},
\]
where $\mathbf{x} \in \mathbb{R}^{1 \times K}$ is the token activation vector 
and $\mathbf{W} \in \mathbb{R}^{K \times N}$ represents the model weights.  
During autoregressive generation, $\mathbf{x}$ updates token by token, 
requiring multiple small GEMV operations executed sequentially.  
Each operation reads the full weight matrix from off-chip memory but performs limited arithmetic, 
making the process inherently \textit{memory-bound}.  
This low arithmetic intensity causes severe underutilization of compute resources on modern accelerators.  
To alleviate this bottleneck, quantization is commonly applied to reduce the storage and bandwidth requirements of $\mathbf{W}$.

\textbf{Vector Quantization.}  
As shown in \Fig{fig:main_arch} (a)~\circlednumberblue{2},  
Vector Quantization (VQ) compresses the weight matrix $\mathbf{W}$ by grouping $d$ consecutive weights along the $K$ dimension into $d$-dimensional vectors.  
Each group of $d$ weights forms a small vector that is replaced by an index referencing a shared weight codebook (WC) $\mathbf{B}$.  
The codebook contains $2^n$ representative $d$-dimensional centroids (\resp{RA-C1}{where $n$ is the index bit-width}), each learned from the distribution of weights using $k$-means clustering~\cite{han2015deep,egiazarian2024extreme}.  
All $K\times N$ weight elements are therefore represented as $V\times N$ indices, where $V = K/d$.  
The weight index (WI) matrix $\mathbf{I}$ stores these compact indices, while the codebook $\mathbf{B}$ stores the learned centroids.

During inference, the quantized model no longer loads the original full-precision weights.  
Instead, each weight vector is reconstructed on demand by fetching its corresponding centroid from the codebook.  
This process replaces bandwidth-intensive FP16 weight loading with lightweight index-based lookups, significantly reducing the storage and memory access cost of LLM layers.

In this study, we adopt $d = 8$ for the vector dimension and $n = 8$ for the index bit-width, 
resulting in a codebook of $2^n = 256$ centroids.  
This corresponds to an average quantization rate of approximately $\frac{n}{d} = 1$ bit per element for one codebook, 
which, however, is too low to preserve model accuracy.  
To achieve higher precision while maintaining compression efficiency, we adopt the additive vector quantization strategy AQLM~\cite{egiazarian2024extreme} 
and use multiple codebooks for hierarchical refinement.  
Specifically, we employ $C$ codebooks to obtain an effective average quantization precision of $q = \frac{C\times n}{d}$ bits.  
\eva{} supports $C = 2$, $3$, and $4$ codebooks, 
corresponding to 2-bit, 3-bit, and 4-bit quantization for LLM decoding.

\textbf{Computation and Memory Inefficiencies.}  
After dequantization, conventional VQ-based methods continue with the GEMV operation as shown in step~\circlednumberblue{1}~\cite{egiazarian2024extreme, tseng2024quip}.  
This offline VQ process significantly reduces memory footprint and data movement.  
However, it does not change the fundamental LLM decoding computation pattern, since each token still performs a GEMV using the reconstructed weights.  
Furthermore, the lookup-based reconstruction introduces irregular memory access patterns that cause frequent bank conflicts and degrade parallel efficiency.  
Compared with 1D $k$-means quantization, VQ demands higher bandwidth and larger codebooks.  
These characteristics increase the memory access complexity of VQ, but they also reveal an inherent structural regularity that can be exploited for more efficient computation.  

\subsection{Recasting GEMV to GEMM}
While conventional VQ increases codebook dimensionality and lookup complexity, its multi-dimensional centroids also expose an opportunity for computation reorganization.  
By exploiting this property, we reformulate the LLM decoding computation so that the original GEMV can be recast as a GEMM operation, thereby improving compute efficiency.  
The detailed process is described below.

\textbf{Dot product between input and centroids.}  
As illustrated in \Fig{fig:main_arch} (b)~\circlednumberblue{3},  
we first reshape the input token vector $\mathbf{x} \in \mathbb{R}^{1\times K}$ into a two-dimensional matrix $\mathbf{X} \in \mathbb{R}^{(K/d)\times d}$ by grouping every $d$ consecutive elements into a vector.  
According to matrix multiplication rules, each $d$-dimensional row vector of $\mathbf{X}$ performs a dot product with the corresponding $d$ elements from each column of the weight matrix $\mathbf{W} \in \mathbb{R}^{K\times N}$.  
After applying VQ, these $d$ elements are replaced by the centroids referenced by the weight index (WI) matrix $\mathbf{I} \in [0,2^n)^{(K/d)\times N}$. Therefore, each input vector interacts with the corresponding centroids across all output channels.  
This means that, regardless of the output dimension $N$, the number of possible weight values for each input vector is bounded by the codebook size $2^n$.  

\textbf{VQ GEMM.}  
This observation enables a key simplification.  
We can directly multiply the reshaped input matrix $\mathbf{X}$ with the codebook $\mathbf{B} \in \mathbb{R}^{d\times 2^n}$ to obtain an intermediate result called the output codebook (OC).  
Each element in the output codebook $\mathbf{O} \in \mathbb{R}^{(K/d)\times 2^n}$ represents the dot product between one input vector and one centroid, which can be reused across multiple output channels.  
This operation effectively converts the token-level GEMV into a compact GEMM between $\mathbf{X}$ and $\mathbf{B}$:
\[
\mathbf{O} = \mathbf{X}\mathbf{B},
\]
where $\mathbf{O}$ is the output codebook.  
This formulation bridges vector quantization and matrix multiplication, paving the way for efficient GEMM-based LLM decoding on modern accelerators.

\textbf{Advantages.}  
This reformulation provides several key benefits for efficient LLM decoding:

\begin{enumerate}
    \item \textit{Regularized memory access.}  
    Converting LLM decoding into GEMM avoids direct lookups on the weight codebook, which are irregular.  
    The number of lookups is reduced, and all memory accesses become regular and coalesced.

    \item \textit{Lower bandwidth demand.}  
    The dot-product operation aggregates every $d$ weights into one value,  
    reducing the lookup bandwidth requirement by a factor of $d$.

    \item \textit{Reduced computation.}  
    For LLMs, the output dimension $N$ often exceeds 4096, while the codebook size is only $2^n = 256$.  
    Conventional GEMV requires $1 \times N \times K$ multiply-accumulate operations,  
    whereas VQ-GEMM needs only $\frac{K}{d} \times 2^n \times d = K \times 2^n$ operations, achieving approximately $\frac{N}{2^n} = 16\times$ fewer computations.

    \item \textit{Higher accelerator utilization.}  
    Recasting LLM decoding into GEMM greatly improves hardware utilization.  
    The matrix dimension in the $M$ direction is no longer fixed at 1,  
    but expanded to $V = \frac{K}{d}$, which is typically greater than 512,  
    enabling efficient use of the accelerator’s matrix--multiplication units.
\end{enumerate}

Together, these advantages allow \eva{} to achieve high throughput with regular memory access, lower bandwidth demand, and significantly improved compute efficiency.

\subsection{Epilogue for Lookup and Reduction}

After the VQ-GEMM operation, we obtain the intermediate result matrix $\mathbf{O}$, as shown in \Fig{fig:main_arch} (b)~\circlednumberblue{4}.  
In the epilogue stage, two operations are required to reconstruct the final output of LLM decoding.

\textbf{1) Lookup using the index matrix.}  
Each output element is selected from $\mathbf{O}$ according to its corresponding index in the weight index matrix $\mathbf{I}$:
\[
\hat{\mathbf{y}} = \text{Lookup}(\mathbf{O}, \mathbf{I}),
\]
where $\hat{\mathbf{y}}$ denotes the intermediate partial sum.  
This lookup is conflict-free because both $\mathbf{I}$ and $\mathbf{O}$ share the same height $V = K/d$.  
Each row of $\mathbf{O}$ is mapped to a dedicated memory bank, enabling parallel access along the $V$ dimension without conflicts.  
Compared with conventional VQ methods, accessing the same number of WI entries does not require additional memory banks.  
However, the bandwidth of each bank is reduced by a factor of $d$, since each access fetches only one FP16 element instead of an $8d$ centroid with 8 FP16s.

\textbf{2) Reduction by accumulation.}  
After lookup, the selected elements are accumulated to form the final output vector.  
This add-only reduction requires no multipliers and incurs minimal hardware cost, allowing the epilogue to run efficiently in parallel with the GEMM pipeline.

\textbf{Advantages.}  
The epilogue is lightweight and hardware-friendly, featuring:  
(i) add-only operations,  
(ii) conflict-free parallel access,  
(iii) reduced bandwidth by a factor of $d$, and  
(iv) a simple, efficient pipeline design.

Overall, this stage completes the LLM decoding process with minimal overhead and provides a clean interface between VQ-GEMM computation and final output reconstruction.

\subsection{Efficient VQ Architecture}  
Finally, \Fig{fig:main_arch} (c) presents the overall \eva{} architecture.  
Thanks to the structured scheduling of the VQ-GEMM operation, the architecture design remains simple, scalable, and highly generalizable across different accelerators.

First, \eva{} adopts a systolic-array-based computation core, which has been widely used in modern AI accelerators~\cite{guo2022ant, guo2023olive, jang2024figna}.  
The proposed VQ-GEMM operation can be directly mapped onto the existing systolic array with minimal modification. 
While VQ requires FP16 arithmetic for accuracy, the LLM prefill stage can maintain accuracy using only INT8 precision~\cite{kim2021bert,xiao2023smoothquant}.
To support both regimes efficiently, we introduce a reconfigurable PE design that performs one FP16 operation or four INT8 operations per cycle, enabling seamless and high-throughput reuse across FP16 VQ decoding and INT8 prefill computation.

Second, a dedicated epilogue unit is introduced to support VQ-specific post-processing, including the lookup and accumulation operations described earlier.  
This unit is optimized for conflict-free access and add-only reduction, ensuring low latency and efficient integration with the systolic array pipeline.

These architectural components and their co-optimization strategies will be discussed in detail in the following sections.
\section{Vector Quantization-based GEMM}
\label{sec:gemm}
This section introduces the VQ-based GEMM design in \eva{}, including the tiling strategy, the mixed-precision GEMM unit, and its architectural integration with the epilogue stage.

\begin{figure}[t]
    \centering
    \includegraphics[width=0.48\textwidth]{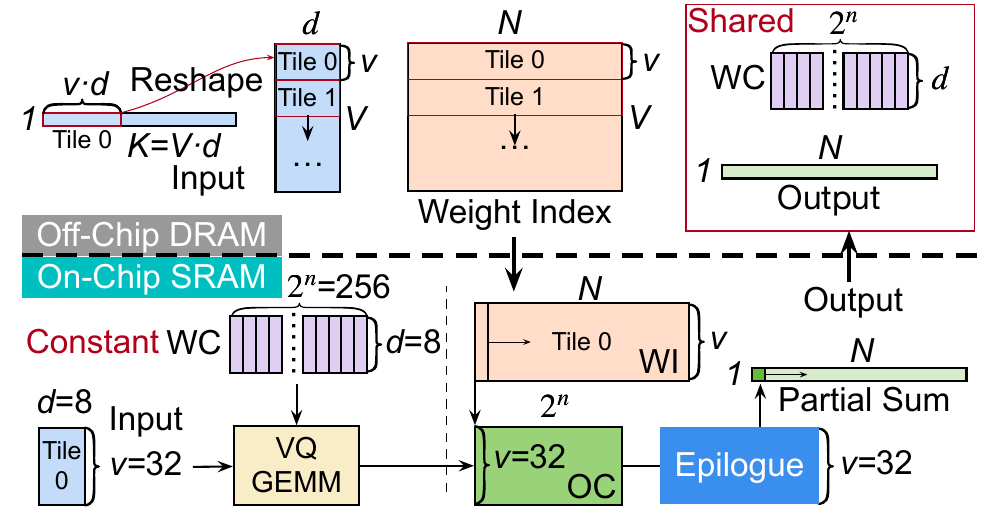}
    \caption{
    Tiling strategy of the VQ-GEMM operation in \eva{}.  
    Input and WI tensors are streamed as tiles from off-chip memory, while the WC and output remain stationary on-chip for reuse.
    }
    \vspace{-5pt}
    \label{fig:tiling}
\end{figure}

\subsection{Tiling Strategy}
Tiling is an essential optimization for GEMM execution because the on-chip memory resources of accelerators are limited.  
As shown in \Fig{fig:tiling}, the tensors involved in the VQ-GEMM computation, including the input, weight index (WI), weight codebook (WC), and output, are pre-arranged in off-chip DRAM before execution.  
Among these tensors, only the input and WI matrices are tiled for streaming access, while the WC and output remain stationary on-chip to maximize reuse.

For the input, the matrix is reshaped and divided into tiles of size $v \times d$, where $v$ is set to 32 in our design.  
Each input tile is loaded into the on-chip buffer for one round of computation.  
Similarly, the WI tensor is accessed by $v$ rows per iteration, corresponding to $v \times N$ elements.  
Since $v \times N$ can be large for modern LLM layers, WI is streamed into the chip to balance throughput and buffer utilization.

In contrast, both the weight codebook (WC) and the output are shared within a layer and remain stationary in on-chip SRAM.  
The WC is read-only throughout the layer, while the output tile is progressively updated after each VQ-GEMM iteration and finally written back to DRAM after all partial sums are accumulated.

The on-chip buffers are organized into two regions:  
(1) one for the input and WC tensors participating in GEMM computation, and  
(2) another for the WI and output tensors, which are consumed and updated during the epilogue stage (\Sec{sec:epilogue}).  

\begin{figure}[t]
    \centering
    \includegraphics[width=0.45\textwidth]{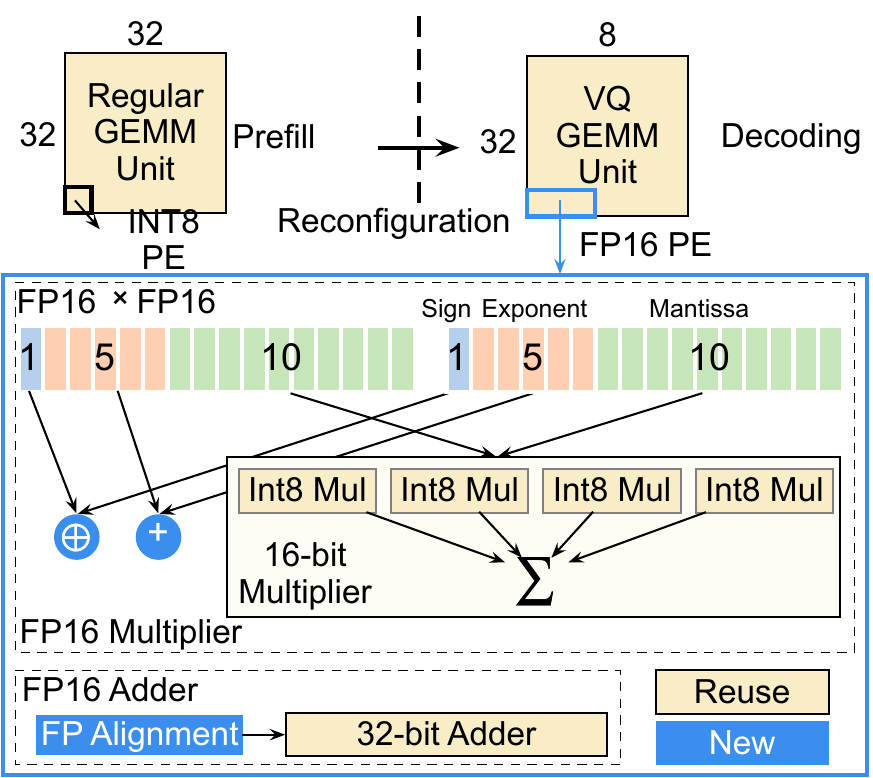}
    \caption{
    Reconfigurable GEMM unit in \eva{}.  
    The mixed-precision PE array supports both INT8 prefill and FP16 VQ-GEMM for LLM decoding through precision reconfiguration.
    }
    \label{fig:gemm}
    \vspace{-5mm}
\end{figure}

\subsection{Mixed-precision Processing Element}
\label{sec:mp}
Although \eva{} is primarily designed for LLM decoding, it must also maintain compatibility with other workloads such as prefill and attention computations.  
For prefill and attention computations, FP16 precision is often unnecessary, and low-precision INT8 arithmetic is sufficient~\cite{kim2021bert,xiao2023smoothquant}.  
Therefore, the base GEMM unit in \eva{} adopts an INT8-based processing element (PE) design, which is widely used in modern accelerators.  
However, LLM decoding relies on weight-only quantization and FP16 computation to maintain accuracy.  
Supporting both modes independently would require a separate FP16 multiplier array, resulting in substantial hardware overhead.  

To address this challenge, \eva{} reuses the existing INT8 multiply-accumulate (MAC) array to support FP16 operations through lightweight reconfiguration, as illustrated in \Fig{fig:gemm}.  
An FP16 number consists of a 1-bit sign, a 5-bit exponent, and a 10-bit mantissa.  
By decomposing each FP16 multiplication into four INT8 multiplications, the PE reconstructs FP16 mantissa multiplication without requiring a dedicated FP16 multiplier.  
A small 6-bit adder and several XOR gates are added to handle exponent and sign computation.  
For FP16 addition, an alignment unit is inserted before accumulation, and the existing INT32 accumulator is reused for the final summation.  

This reconfiguration enables FP16 support with minimal hardware modification.  
A conventional $32\times32$ INT8 array can be dynamically reconfigured into a $32\times8$ array for FP16 computation during LLM decoding.  
This configuration perfectly matches the tiling parameters ($v=32$, $d=8$) used in the VQ-GEMM operation, allowing the GEMM unit to achieve optimal utilization and performance.  
The result is a unified GEMM unit capable of executing both INT8 and FP16 operations with high area efficiency and utilization, enabling \eva{} to flexibly support both prefill and LLM decoding workloads.

\begin{figure}[t]
    \centering
    \includegraphics[width=0.48\textwidth]{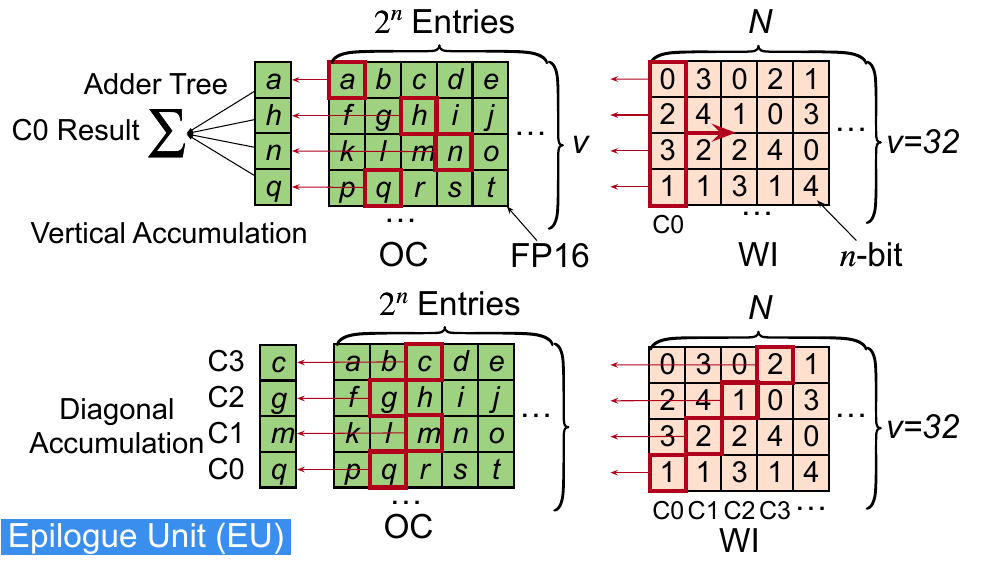}
\caption{\resp{RF-Q3}{
Epilogue unit (EU) for conflict-free lookup in \eva{}, supporting vertical adder-tree reduction and diagonal accumulation across outputs (C0--C3) with output-level parallelism.
}}
    \label{fig:epilogue}
    \vspace{-5mm}
\end{figure}

\section{Epilogue Unit and Pipeline}
\label{sec:epilogue}

\subsection{Epilogue Unit}
The epilogue unit (EU) in \eva{} is designed to be lightweight, parallel, and highly efficient.  
As shown in \Fig{fig:epilogue}, it performs two core operations: conflict-free lookup and add-only accumulation.  

\textbf{Basic Units.}
\resp{RF-Q3}{
Each EU performs a fast $n$-bit index lookup followed by an add-only reduction.
For each tile, $v=32$ weight indices (WIs) are read in parallel and used to fetch the corresponding FP16 values from the output codebook (OC).
The retrieved values are accumulated using one of two schemes:
(1) a vertical 32-input adder-tree reduction for single-codebook execution, or
(2) a diagonal accumulation scheme that enables output-level parallel reduction across multiple codebooks (e.g., C0--C3).
Both designs maintain conflict-free access and high utilization while avoiding multipliers in the epilogue stage.
}

\textbf{EU Efficiency.}  
Although each EU contains only 32 adders, each accumulates $d=8$ dot-product results rather than a single scalar.  
Thus, the 32 adders collectively process $\frac{32\times d}{C}=128$ weight elements and their corresponding dot products, equivalent to conventional VQ decoding.
In contrast, traditional designs require conflict-prone memory accesses and costly MAC operations.  
By using this add-only scheme, \eva{} achieves the same functionality with lower computation and hardware cost, without sacrificing arithmetic precision.

\begin{figure}[t]
    \centering
    \includegraphics[width=0.45\textwidth]{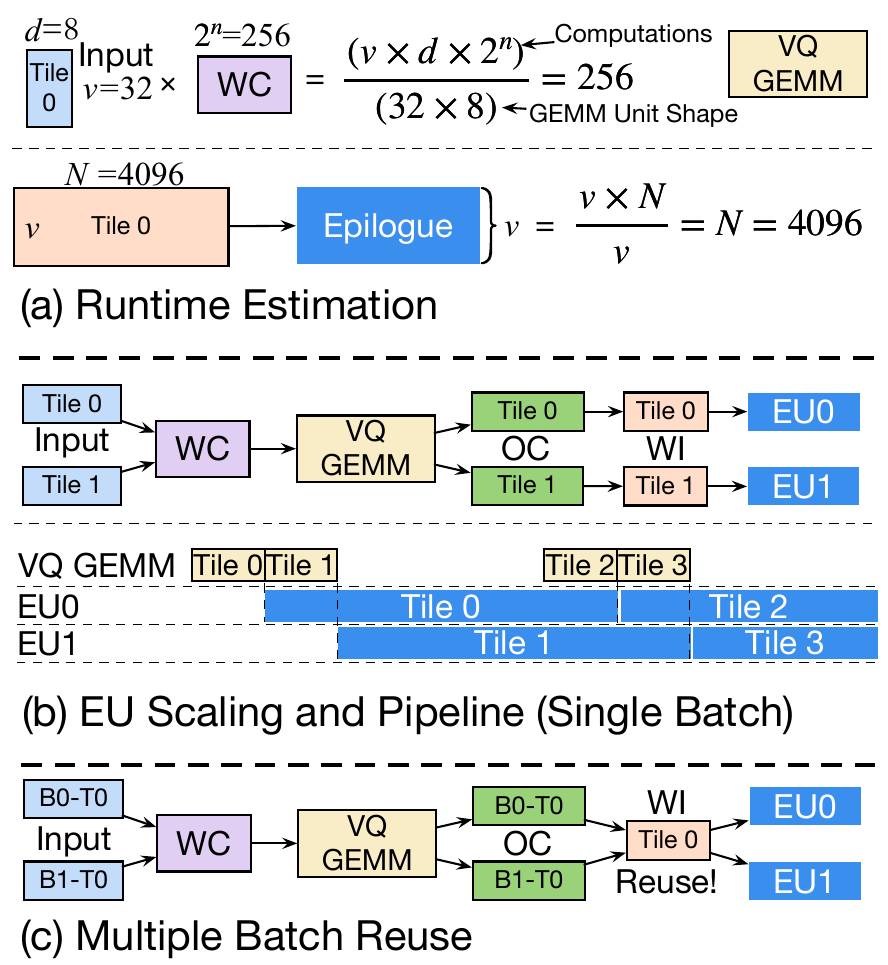}
    \caption{
    Execution scheduling of \eva{}.  
    (a) Runtime estimation showing that the GEMM stage is not the bottleneck.  
    (b) EU scaling pipeline demonstrating GEMM--Epilogue overlap.  
    (c) Multi-batch reuse, where multiple requests share the same weight tiles to reduce bandwidth cost.
    }
    \label{fig:scheduling}
    \vspace{-5mm}
\end{figure}

\subsection{EU Scaling}
\Fig{fig:scheduling} (a) illustrates the runtime estimation and pipeline organization of \eva{}.  
The execution latency of VQ-GEMM on a $32\times8$ GEMM unit depends only on the tiling parameters $v$, $d$, and $2^n$, and is independent of the original matrix width $N$.  
For instance, when $v=32$, $d=8$, and $2^n=256$, one VQ-GEMM operation requires only 256 cycles, whereas the EU ($N=4096$) would take 4096 cycles.  
In this configuration, GEMM is not the bottleneck; instead, the critical path resides in the epilogue unit (EU).  

As shown in \Fig{fig:scheduling} (b), the system overlaps VQ-GEMM computation with epilogue processing to maximize hardware utilization and throughput.  
The output tiles produced by the GEMM unit are directly consumed by the EUs without off-chip transfers, reducing latency and avoiding bandwidth contention.  
This concurrent scheduling allows the EU to sustain near-peak utilization across all computation stages.

Consequently, GEMM units become partially idle as the bottleneck shifts from multiplication to addition.  
Since the EU consists only of adders and requires much lower bandwidth, its scaling cost is minimal.  
Therefore, overall performance can be improved by simply increasing the number of EUs to better match the GEMM output rate.  
However, the number of EUs cannot grow indefinitely and must be jointly optimized with respect to model size, VQ configuration, GEMM throughput, and memory bandwidth.  
We conduct a design-space exploration (DSE) in \Sec{sec:dse} to analyze these trade-offs.

\subsection{Batch Scaling}
\eva{} also efficiently supports multi-batch execution, enabling system-level optimizations such as continuous batching.  
As shown in \Fig{fig:scheduling} (c), when handling two requests, each with its own Tile~0, the VQ-GEMM stage operates identically to the single-request case.  
At the EU stage, the Tile~0 results from both requests can reuse the same weight tile, significantly reducing bandwidth consumption.  
This reuse capability improves overall throughput and efficiency for multi-batch workloads.  
We further evaluate the effect of batch scaling on \eva{}'s performance in \Sec{sec:eval}.

\section{Evaluation}
\subsection{Methodology}
\textbf{Accelerator Baselines.}
We compare \eva{} with four baseline architectures: Systolic Array (SA)~\cite{jouppi2017datacenter}, ANT~\cite{guo2022ant}, FIGNA~\cite{jang2024figna}, and FIGLUT~\cite{park2025figlut}. The Systolic Array~\cite{jouppi2017datacenter} optimizes matrix multiplication by enhancing data reuse and reducing external memory bandwidth. ANT~\cite{guo2022ant} introduces a fixed-length data type and an adaptive hardware framework for low-bit quantization. FIGNA~\cite{jang2024figna} converts floating-point activations into integers using pre-alignment, allowing for efficient integer-only computation. FIGLUT~\cite{park2025figlut} replaces FP-INT GEMM with a look-up table architecture that retrieves precomputed partial sums of activations based on weight patterns.

\textbf{Accelerator Implementation.} 
We develop our own simulator based on the validated open-source simulator, ANT~\cite{guo2022ant}. All baseline accelerators are integrated into the simulator to compare their performance with \eva{}. We implement the hardware architectures for \eva{} and all baseline models in Verilog HDL. To ensure a fair comparison, we synthesize all designs using Cadence Genus with the TSMC 28nm technology library, targeting a clock frequency of 500MHz to measure hardware area and power. Additionally, we simulate the SRAM buffer area and power using Cacti 7.0~\cite{balasubramonian2017cacti}, which is also based on the 28nm process. Finally, the power consumption of DRAM is simulated based on \resp{RE-C4}{DRAMsim3}~\cite{li2020dramsim3}.

\textbf{Benchmark.} We evaluate \eva{} with baselines using the LLaMA 1, 2, and 3 models~\cite{touvron2023llama1,touvron2023llama2,llama2024llama3}, as well as two advanced Mixture-of-Experts (MoE) models: Mixtral-8x7B~\cite{jiang2024mixtral} and Qwen3-30B-A3B~\cite{yang2025qwen3}. 
We employ two types of datasets for our benchmarking. First, we use a synthetic dataset with a fixed input length (M) to evaluate model performance during the decoding phase. Second, we assess end-to-end performance in real-world LLM inference scenarios using the Dolly~\cite{DatabricksBlog2023DollyV2}, Arxiv Summarization~\cite{cohan2018discourse}, and GSM8K~\cite{cobbe2021training} datasets. To simplify our experiments, we run the first Transformer block of each model.

\begin{table}[t]
  \centering
  \small
  \renewcommand{\arraystretch}{1.2}
  \caption{\resp{CQ2}{Normalized \eva{}'s Latency on LLaMA-2-7B Across Different VQ Configurations.}}
  \resizebox{\columnwidth}{!}{%
    \begin{tabular}{l | c c c c c |c | r}
      \toprule
      \textbf{Algorithm} & $\mathbf{d}$ & $\mathbf{n} \ (2^n)$ & $\mathbf{C}$ & $\mathbf{q}$ & ${N}$ & \textbf{PE:EU} & \textbf{Norm. Latency} \\
      \midrule
      AQLM 2$\times$8 & 8 & 8 (256) & 2 & 2 & 4096 & $1:16$ & 1.00$\times$ \\
      AQLM 3$\times$8 & 8 & 8 (256) & \textbf{3} & \textbf{3} & 4096 & $1:16$ & \textbf{1.49$\times$} \\
      AQLM 2$\times$12 & 8 & \textbf{12 (4096)} & 2 & \textbf{3} & 4096 & $1:1$ & \textbf{2.96$\times$} \\
      AQLM 4$\times$8 & 8 & 8 (256) & \textbf{4} & \textbf{4} & 4096 & $1:16$ & \textbf{1.98$\times$} \\
      AQLM 1$\times$16 & 8 & \textbf{16 (65536)} & \textbf{1} & 2 & 4096 & $16:1$ & \textbf{22.86$\times$} \\
      GPTVQ-4D & \textbf{4} & 8 (256) & \textbf{1} & 2 & \textbf{256} & $1:1$ & \textbf{4.17$\times$} \\
      Hypothesized & \textbf{4} & 8 (256) & \textbf{1} & 2 & \textbf{4096} & $1:16$ & \textbf{1.00$\times$} \\
      \bottomrule
    \end{tabular}%
  }
  \label{tab:vq_gains}
    \vspace{-5pt}
\end{table}

\begin{figure}[tb]
    \centering
    \includegraphics[width=0.48\textwidth]{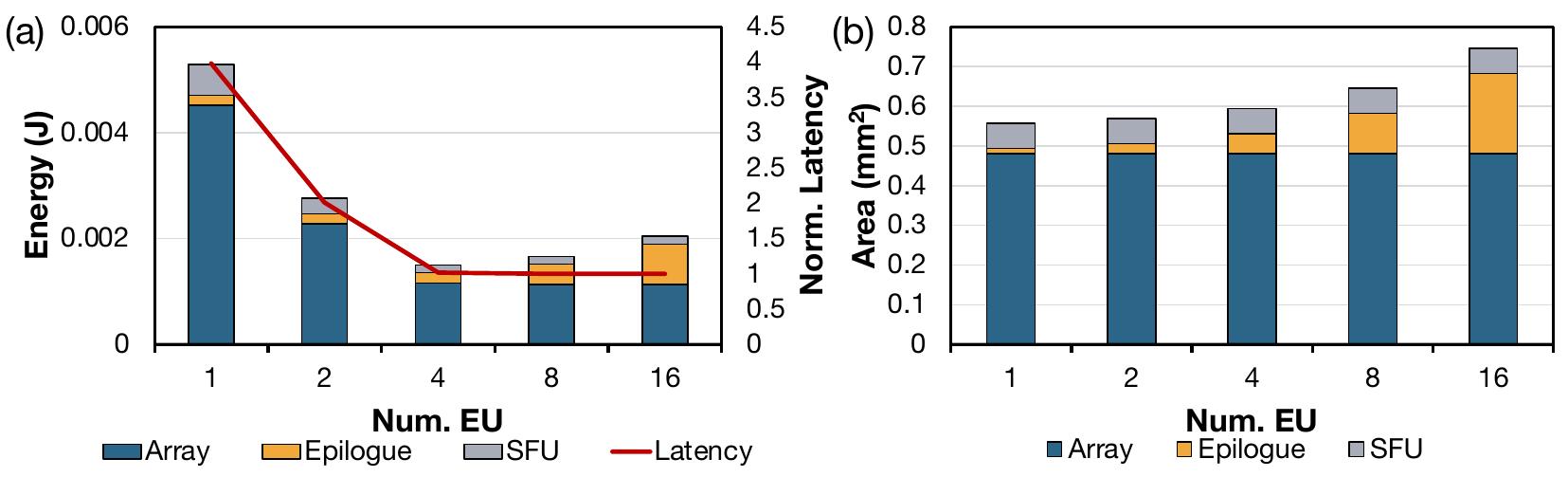}
    \caption{
    Design space exploration of \eva{}.  
    (a) Number of Epilogue Units with latency and energy. (b) Number of Epilogue Units with area.
    }
    \vspace{-5mm}
    \label{fig:dse}
\end{figure}

\subsection{Design Space Exploration}
\label{sec:dse}

We conduct design space exploration on three key vector quantization (VQ) parameters \( n \), \( C \), and \( d \) (defined in \Tbl{tab:variables}) and the number of Epilogue Units (EUs) to identify the optimal \eva{} configuration.


\textbf{VQ parameters.}
\Tbl{tab:vq_gains} shows how \eva{}'s latency varies across different VQ configurations. 
Here, ${N}$ denotes the minimum number of output channels sharing the same codebook; for AQLM, this corresponds to the linear-layer output dimension (${N} \ge 4096$).
As shown in the table, when $2^n < {N}$, latency is approximately proportional to the effective bitwidth $q=\frac{nC}{d}$; thus configurations with the same $q$ exhibit similar performance.
After VQ reformulation, computation splits into PE-side multiplications $K\times2^n$ and EU-side accumulations $K\times {N}$, yielding the ratio $\text{PE:EU}=\frac{2^n}{N}$.
For example, when $n=12$ and $N=4096$, we have $2^n={N}$ (PE:EU = 1:1). In this balanced regime, the systolic array latency exceeds the Epilogue latency, causing \eva{} to become PE-bound and reducing its acceleration benefit. When $2^n > {N}$, this effect becomes more pronounced; centroid under-utilization introduces ``spurious'' multiplications (presented in the \Sec{sec:dis}), further increasing latency.
Our design-space exploration confirms that $2^n=256$ ($n=8$) provides the most practical trade-off between compute efficiency and memory cost.

\textbf{Number of Epilogue Units.} As shown in \Fig{fig:dse} (a) and \Fig{fig:dse} (b), we fix the DRAM bandwidth at 64~GB/s and evaluate the impact of EU count on latency, energy, and area. Increasing the number of EUs initially reduces both latency and energy, but beyond four EUs, latency no longer decreases. This is because in each cycle the four EUs can process \(4\times v = 128\) weight indexes, which fully utilizes the 64 GB/s DRAM bandwidth at a clock frequency of 500 MHz. Further increasing EU only raises energy consumption. Since each EU only consists of adders, its area overhead remains negligible (3.5\%) compared to the PE array when the number of EUs is small.

\textbf{Architecture configuration.} Our exploration indicates that during the decoding stage, the best trade-off is achieved by matching the number of EUs with memory bandwidth. However, higher bandwidth can lead to lower utilization during the prefill phase and increase DRAM power consumption. Additionally, increasing the number of EUs results in a rise in both the architecture's area and buffer size. Therefore, we have chosen to implement four EUs with a memory bandwidth of 64 GB/s. A complete summary of the architectural configuration of \eva{} is presented in \Tbl{tab:config}.

\begin{table}[tb]
    \vspace{-5pt}
    \caption{\eva{} Architectural Configuration}
    \vspace{-8pt}
        \centering
    \resizebox{\columnwidth}{!}{%
        \begin{tabular}{l|l}
            \toprule
            PE Array  & $32\times 32$ INT8, Weight Stationary;\\
            Mixed-precision   & $32\times 8$ FP16, Input Stationary.\\
            \midrule
            Epilogue Unit  & $4\times$ 32-input Adder Tree.\\ \midrule
            Decoding Stage Tiling & $m=M$; $k=\frac{4\times v\times d}{M}$; $n=N$. \\ \midrule
            Prefill Stage Tiling & $m=1024$; $k=32$; $n=1024$. \\ \midrule
            Quantization: $q$-bit & $d=8$; $n=8$; $C=q$.\\ \midrule
            \multirow{2}{*}{Buffer Size: 528KB} & 16KB Weight Codebook; 256KB Weight; 32KB Input; \\ 
                         &  192KB Output Codebook; 32KB Output. \\ \midrule
            DDR4 DRAM & 8Gb $\times$ 8 2133R; 16GB/s per channel; 4 channels 64GB/s. \\
            \bottomrule
        \end{tabular}
        }
        \label{tab:config}
        \vspace{-5pt}
\end{table}

\begin{table}[tb]
\vspace{-5pt}
  \centering
  \small
  \renewcommand{\arraystretch}{1.2}
  \caption[]{The Perplexity on Wikitext data.}
  \resizebox{1\columnwidth}{!}{
  \scriptsize
    \begin{tabular}{l|cc|ccc|cc|c}
      \toprule
      \textbf{Arch.} & {SA} & {ANT} & {FIGNA} & {FIGLUT} & \textbf{\eva{}} & {FIGNA} & \textbf{\eva{}} & {FP16} \\
      \midrule
      \textbf{Algo.} & QSERVE & ANT & AWQ & BCQ & AQLM & AWQ & AQLM & FP16 \\
      \midrule
      \textbf{FC Act.} & INT8 & 8bit & FP16 & FP16 & FP16 & FP16 & FP16 & FP16 \\
      \textbf{FC Wgt.} & INT8 & 8bit & INT4 & 4bit & 4bit & INT2 & 2bit & FP16 \\ \hline
      \textbf{L-2 7B} & 5.56 & 5.58 & 5.60 & 5.58 & \textbf{5.43} & 2.2e5 & \textbf{6.69} & 5.12 \\
      \textbf{L-2 13B} & 4.95 & 5.20 & 4.97 & 4.96 & \textbf{4.76} & 1.2e5 & \textbf{5.63} & 4.57 \\
      \bottomrule
    \end{tabular}
  }
  \label{tbl:acc_wiki}
\end{table}

\begin{table}[tb]
  \centering
  \small
  \renewcommand{\arraystretch}{1.2}
  \caption{Accuracy(\%) Comparison of LLaMA-2-7B on downstream benchmark.}
  \resizebox{\columnwidth}{!}{
  \scriptsize
    \begin{tabular}{l |c |c c c c c}
      \toprule
      \textbf{Method} & \textbf{Bits} & \textbf{PIQA} & \textbf{COPA} & \textbf{ARC-E} & \textbf{ARC-C} & \textbf{Winogrande} \\
      \midrule
      FP16 & 16 & 78.1 & 87.0 & 76.4 & 43.5 & 69.1 \\ 
      \midrule
      \textbf{LLM.265 FB} & 4 & 78.8 & -- & 73.83 & 44.1 & 69.7 \\
      \textbf{LLM.265 VB} & 4 & 78.9 & -- & 73.82 & 45 & 69.7 \\
      \textbf{\eva{}} & 4 & 77.0 & 85.0 & 74.1 & 41.0 & 68.5 \\
      \midrule
      \textbf{LLM.265 FB} & 2 & 54.3 & 68.5 & 29.76 & 30.5 & 51.8 \\
      \textbf{LLM.265 VB} & 2 & 56.7 & 68.9 & 34.52 & 31.1 & 52.3 \\
      \textbf{\eva{}} & 2 & \textbf{75.9} & \textbf{84.0} & \textbf{71.7} & \textbf{38.6} & \textbf{68.2} \\
      \bottomrule
    \end{tabular}
  }
  \label{tbl:acc_bench}
\end{table}

\begin{table}[tb]
  \centering
  \small
  \renewcommand{\arraystretch}{1.2}
  \caption{\resp{RA-Q2}{Accuracy (\%) Comparison of Mixture of Experts (MoE) Models.}}
  \resizebox{\columnwidth}{!}{
  \scriptsize
    \begin{tabular}{l |c |c c c c c}
      \toprule
      \textbf{Method} & \textbf{Bits} & \textbf{ARC-C} & \textbf{ARC-E} & \textbf{PIQA} & \textbf{BoolQ} & \textbf{Winogrande} \\
      \midrule
      \multicolumn{7}{c}{\textbf{Mixtral-8x7B}} \\
      \midrule
      FP16 & 16 & 59.81 & 83.54 & 83.73 & 85.26 & 76.56 \\ 
      \midrule
      \textbf{AWQ} & 4 & \textbf{58.87} & 82.58 & 82.97 & 84.34 & \textbf{76.24} \\
      \textbf{AQLM-2$\times$16} & 4 & 54.61 & 83.12 & 81.99 & -- & 74.82 \\
      \textbf{AQLM-4$\times$8} & 4 & \textbf{58.87} & \textbf{83.38} & \textbf{83.51} & \textbf{85.02} & 76.01 \\
      \midrule
      \textbf{GPTQ} & 2 & 27.30 & 35.44 & 59.79 & 52.08 & 50.83 \\
      \textbf{GPTVQ-4D} & 2 & 46.42 & 65.57 & 78.13 & 78.59 & 71.11 \\
      \textbf{AQLM-1$\times$16} & 2 & 47.93 & 77.68 & 80.43 & -- & \textbf{75.93} \\
      \textbf{AQLM-2$\times$8} & 2 & \textbf{50.43} & \textbf{78.24} & \textbf{80.69} & \textbf{81.28} & 71.74 \\
      \midrule
      \multicolumn{7}{c}{\textbf{Qwen3-30B-A3B}} \\
      \midrule
      FP16 & 16 & 62.80 & 83.75 & 80.36 & 88.47 & 73.88 \\
      \midrule
      \textbf{AWQ} & 4 & \textbf{61.43} & 82.66 & 80.90 & 88.87 & 73.01 \\
      \textbf{AQLM-4$\times$8} & 4 & 61.26 & \textbf{82.95} & \textbf{80.96} & \textbf{89.02} & \textbf{73.72} \\
      \midrule
      \textbf{AQLM-2$\times$8} & 2 & \textbf{54.27} & \textbf{76.85} & \textbf{76.93} & \textbf{86.94} & \textbf{69.69} \\
      \bottomrule
    \end{tabular}
  }
  \label{tbl:acc_moe}
  \vspace{-5mm}
\end{table}

\subsection{Model Performance}
\textbf{Comparison of \eva{} and the baselines.} We evaluate the LLaMA-2-7B (L-2 7B) and LLaMA-2-13B (L-2 13B) on the WikiText-2 dataset~\cite{merity2016pointer}, comparing \eva{} with accelerator baselines using their respective state-of-the-art (SOTA) quantization algorithms. 

As shown in \Tbl{tbl:acc_wiki}, systolic arrays adopt INT8 multiply-accumulate (MAC) units with QSERVE~\cite{lin2024qserve}, ANT applies its dedicated 8-bit format, FIGNA supports FP16 activation and INT4 weight for AWQ~\cite{lin2024awq}, FIGLUT uses binary-coding quantization (BCQ) with ShiftAddLLM~\cite{you2024shiftaddllm}, and \eva{} employs AQLM~\cite{egiazarian2024extreme, malinovskii2024pvtuning} with parameters $d=8, n=8$. All methods maintain Attention layers in FP16 precision.

\Tbl{tbl:acc_wiki} presents the perplexity results, where lower perplexity indicates better performance. At higher quantization precision (e.g., 8-bit or 4-bit weight-only), all algorithms achieve comparable accuracy with minimal degradation from FP16, while \eva{} slightly outperforms others. When precision decreases to 2-bit, weight-only quantization collapses, but the VQ-based method maintains competitive accuracy due to its higher-dimensional clustering.

We further compare \eva{} with the video-codec-based LLM.265~\cite{xu2025llm} framework on downstream benchmarks, including PIQA~\cite{bisk2020piqa}, COPA~\cite{roemmele2011choice}, ARC-Easy, ARC-Challenge~\cite{clark2018think}, and Winogrande~\cite{sakaguchi2021winogrande}. As shown in \Tbl{tbl:acc_bench}, at higher bitwidths, all methods perform similarly, while only VQ-based \eva{} retains competitive accuracy at 2-bit precision. On average, \eva{} improves accuracy by 19 percentage points over LLM.265 VB at 2-bit across all benchmarks.

\textbf{\eva{}'s accuracy trade-off.} We compare vector quantization algorithms used in the \eva{} (AQLM~\cite{egiazarian2024extreme} and GPTVQ~\cite{van2024gptvq}), with weight-only methods on Mixture of Experts (MoE) models (Mixtral-8x7B~\cite{jiang2024mixtral} and Qwen3-30B-A3B~\cite{yang2025qwen3}). We evaluate the accuracy on various downstream datasets, including ARC-Easy, ARC-Challenge~\cite{clark2018think}, PIQA~\cite{bisk2020piqa}, BoolQ~\cite{clark2019boolq}, and Winogrande~\cite{sakaguchi2021winogrande}.

As shown in \Tbl{tbl:acc_moe}, the 4-bit VQ algorithm achieves nearly lossless compression, with an average accuracy decrease of less than 0.5\%. At 2 bits, the AQLM $2\times 8$ configuration shows a 5.3 percentage-point accuracy drop on Mixtral-8x7B, outperforming the non-analytic GPTQ method (32.7\% drop). We find no significant accuracy improvements with larger codebook sizes ($n$) or fewer output channels ${N}$ at the same bit width. In particular, increasing the number of codebooks $C$ can improve accuracy while keeping $n$ small. \Tbl{tab:vq_gains} and \Tbl{tbl:acc_moe} show that the AQLM $1\times16$ configuration with lower bit width performs worse than the AQLM $4\times 8$ configuration with higher bit width on both efficiency and accuracy. Thus, using the VQ configuration with $n=8$ for the \eva{} evaluation is justified.

Overall, \Tbl{tbl:acc_wiki}, \Tbl{tbl:acc_bench}, and \Tbl{tbl:acc_moe} highlight the robustness of VQ-based quantization at low precision. Importantly, \eva{} is decoupled from specific quantization algorithms: it does not rely on any particular method and can benefit from future improvements, such as fine-tuning~\cite{malinovskii2024pvtuning} or other emerging optimizations~\cite{tseng2024quip,zhang2405kv,li2025commvq}.

\begin{table}[tb]
  \centering
  \vspace{-6pt}
  \caption{The area, power, throughput, and efficiency comparison of \eva{} and baseline accelerators (28nm, 500\,MHz).}
  \renewcommand{\arraystretch}{1.2}
  \resizebox{0.48\textwidth}{!}{%
  \begin{tabular}{l|ccccc}
    \toprule
    \textbf{Architecture} & \textbf{SA} & \textbf{ANT} & \textbf{FIGNA} & \textbf{FIGLUT} & \textbf{\eva{}} \\
    \midrule
    Area ($\mathrm{mm}^2$) & 1.256 & 1.472 & 1.211 & 1.582 & 1.414 \\
    \midrule
    On-chip Power (W) & 1.647 & 2.741 & 2.602 & 4.037 & 3.117 \\
    \midrule
    Throughput & 15.75 & 15.28 & 14.84 & 44.49 & \textbf{498.49} \\
    (GOPs) & (1.00$\times$) & (0.97$\times$) & (0.94$\times$) & (2.82$\times$) & (31.64$\times$) \\
    \midrule
    Area Efficiency & 12.54 & 10.38 & 12.25 & 28.12 & \textbf{352.54} \\
    (GOPs/mm$^2$) & (1.00$\times$) & (0.83$\times$) & (0.98$\times$) & (2.24$\times$) & (28.10$\times$) \\
    \midrule
    Energy Efficiency & 9.56 & 5.58 & 5.70 & 11.02 & \textbf{159.94} \\
    (GOPs/W) & (1.00$\times$) & (0.58$\times$) & (0.60$\times$) & (1.15$\times$) & (16.72$\times$) \\
    \midrule
    \multicolumn{6}{c}{Buffer: 528\,KB; DRAM Bandwidth: 64\,GB/s.} \\
    \bottomrule
  \end{tabular}%
  }
  \label{tab:area_power}
\end{table}

\begin{figure}[tb]
    \centering
    \includegraphics[width=0.48\textwidth]{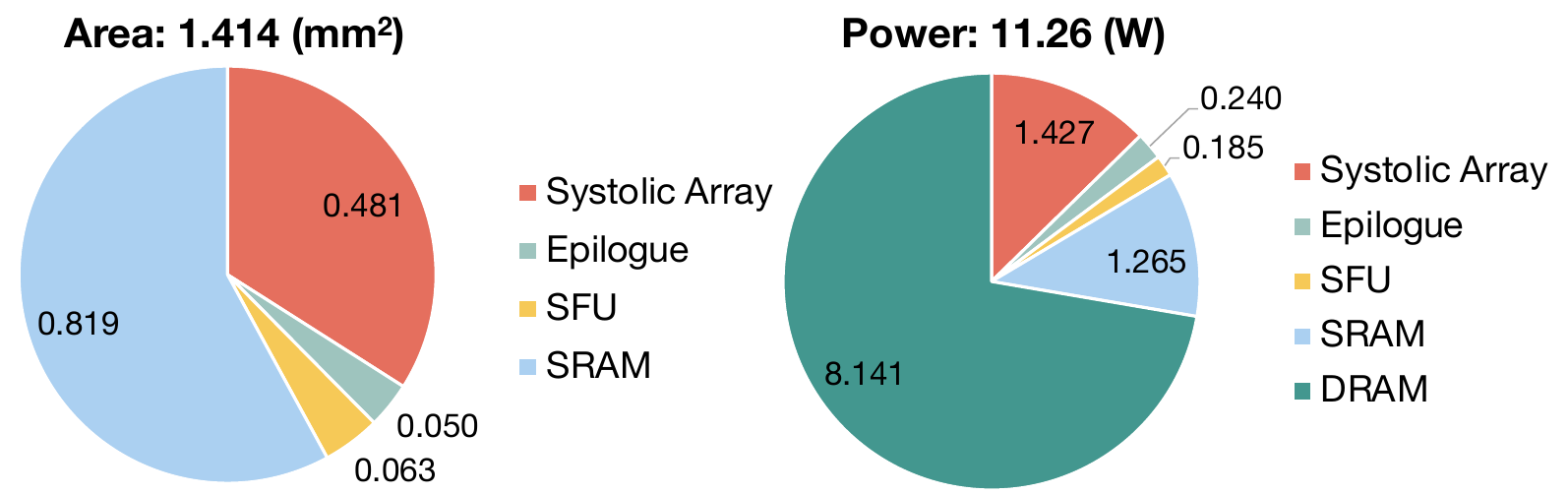}
    \caption{
    \eva{} area and power breakdown.
    }
    \label{fig:breakdown}
    \vspace{-10pt}
\end{figure}

\subsection{Area and Power Comparison}

As shown in \Tbl{tab:area_power}, we compare the area and power of \eva{} and the baseline architectures. For a fair comparison, all designs are configured with the same minimum number of PEs. For \eva{}, compared to the INT8 systolic array, the increases in area and power are primarily due to its mixed-precision support for FP16 computation during the decoding stage. Although \eva{} does not have the smallest area or power consumption, it achieves significantly higher throughput and demonstrates superior energy and area efficiency using 2-bit VQ, highlighting the effectiveness of the proposed design.

\textbf{Breakdown.} \Fig{fig:breakdown} illustrates the area and power breakdown of \eva{}. On-chip SRAM occupies the largest area and has the second-highest power consumption, mainly due to double buffering and the frequent accesses to the output-codebook buffer. Off-chip DRAM accounts for the highest power consumption, which is consistent with the memory-bound nature of the decoding phase. Within the compute core, the PE array occupies the main area and power consumption, while the Epilogue Units introduce only a small overhead.

\begin{figure*}[t]
    \centering
    \includegraphics[width=\linewidth]{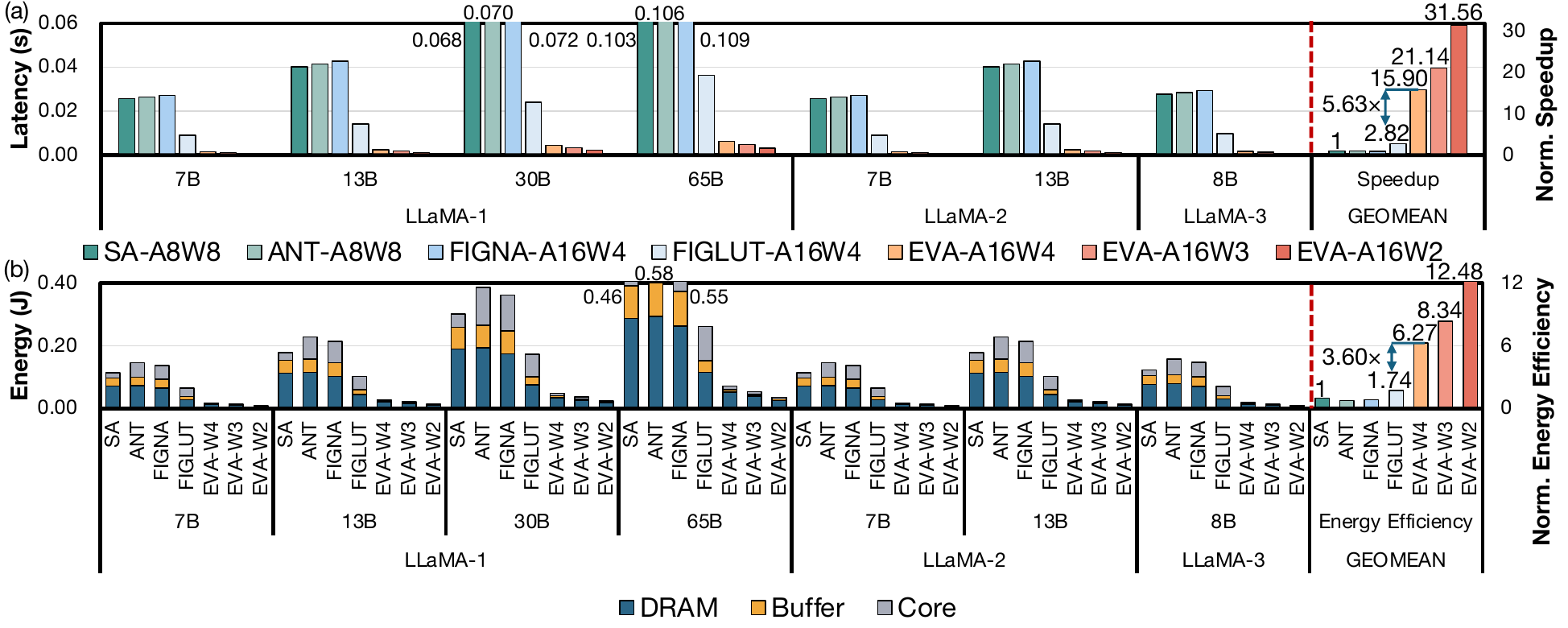}
    \caption{
    Latency and energy consumption of the \eva{} and baseline accelerators on the fully connected layer with batch size=1 during the decoding phase of the LLaMA models. Method-A$n$W$m$ denotes n-bit activation and m-bit weight.
    }
    \vspace{-5mm}
    \label{fig:se}
\end{figure*}

\subsection{\eva{} Performance and Energy}
\label{sec:eval}

\textbf{Latency and energy on single-batch decoding.} We evaluate \eva{} and the baselines on the fully connected (FC) layers of the LLaMA models during the decoding phase with a batch size of 1. As shown in \Fig{fig:se} (a), all baselines suffer from low array utilization because decoding is executed as GEMV, which leads to long latency. On the $32\times32$ systolic array, only one lane is effectively active when batch size $=1$. The utilization rates of ANT and FIGNA are further reduced due to the increased pipeline fill and drain overhead. As a LUT-based method, FIGLUT attains higher utilization by using 4-input LUTs and broadcasting their outputs to multiple PEs. However, its utilization rate remains relatively low (4.34\%). 

In contrast, \eva{}-A16W2 achieves higher hardware utilization by transforming GEMV into GEMM with a conflict-free output LUT. Therefore, \eva{}-A16W2 achieves speedups of 31.56$\times$, 32.53$\times$, 33.50$\times$, and 11.17$\times$ over SA, ANT, FIGNA, and FIGLUT, respectively. Furthermore, \eva{}-A16W2 delivers 1.99$\times$ and 1.49$\times$ speedups over \eva{}-A16W4 and \eva{}-A16W3, respectively. 
This improvement arises from the fact that lower-bit VQ uses fewer codebooks~\cite{egiazarian2024extreme,tseng2024quip}, which proportionally reduces the associated computational cost.

\Fig{fig:se} (b) reports the energy consumption. For all architectures, DRAM access dominates the total energy cost. Compared with SA, ANT, FIGNA, and FIGLUT, \eva{}-A16W2 achieves 12.48$\times$, 15.96$\times$, 14.96$\times$, and 7.17$\times$ higher energy efficiency, respectively. \eva{}-A16W2 further improves energy efficiency by 1.99$\times$ and 1.50$\times$ over \eva{}-A16W4 and \eva{}-A16W3, respectively.

\begin{figure}[tb]
    \centering
    \includegraphics[width=0.48\textwidth]{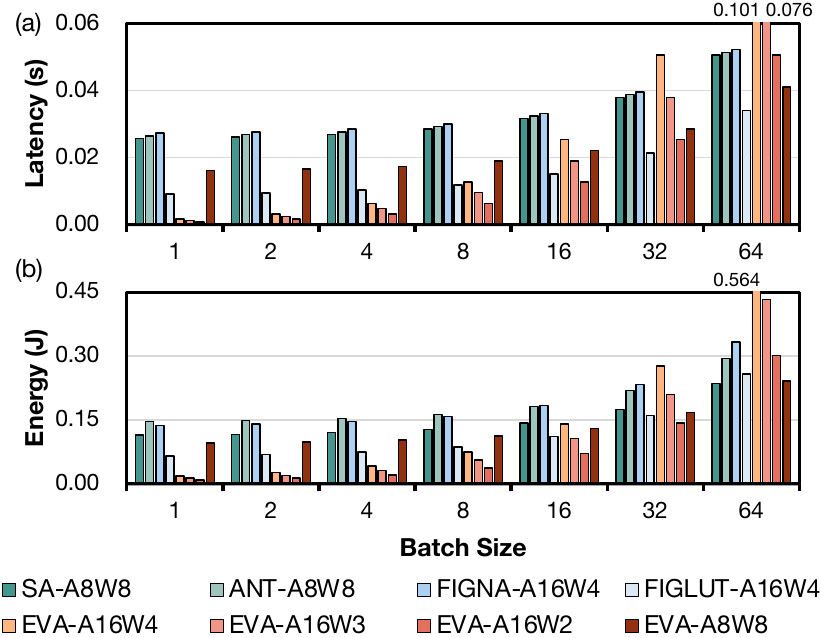}
    \caption{Effect of batch scaling on LLaMA-2-7B.}
    \label{fig:batch}
    \vspace{-10pt}
\end{figure}

\textbf{Effect of batch scaling.} We assess the effect of multi-batch execution during the decoding phase on \eva{} and the baseline accelerators using the LLaMA-2-7B model. Here, \eva{}-A8W8 refers to the INT8 computation method described in \Sec{sec:mp}. 

As shown in \Fig{fig:batch} (a), the latency for SA, ANT, FIGNA, FIGLUT, and \eva{}-A8W8 increases slowly when the batch size is less than 8. This is because their hardware utilization remains low at smaller batch sizes, so the extra computation is partly hidden by improved array utilization.
In contrast, \eva{}-A16W4, \eva{}-A16W3, and \eva{}-A16W2 have already achieved high utilization when the batch size is small, leading to a nearly linear increase in latency as the batch size grows.

\Fig{fig:batch} (b) shows how energy consumption varies with batch size. The energy consumption of FIGNA increases faster than that of ANT. This is primarily due to increased overhead in DRAM accesses. We observe that when the batch size is 1, DRAM access accounts for 48.13\% of the total energy in FIGNA and 50.27\% in ANT. When the batch size becomes 64, these shares increase to 58.28\% and 51.75\%, respectively. This difference arises because FIGNA uses FP16 activations, which require about twice the memory bandwidth of ANT's 8-bit activations. The same effect is observed between FIGLUT and \eva{}-A8W8.

When the batch size is larger than 32, the latency and energy of \eva{}-A16W2 become higher than \eva{}-A8W8. The same trend holds for \eva{}-A16W4 and \eva{}-A16W3. As the batch size continues to increase, the workload changes from GEMV-style to GEMM-style computation, and the array utilization of all non-VQ architectures approaches nearly 100\%. In this regime, the latencies of different architectures become similar, and the energy consumption mainly reflects their power.

\begin{table}[tb]
  \vspace{-0mm}
  \centering
  \small
  \renewcommand{\arraystretch}{1}
  \caption{\resp{RF-Q2}{Input and Output Lengths across Models and Datasets.}}
  \resizebox{0.98\linewidth}{!}{
  \begin{tabular}{l | c | c c | c c}
    \toprule
    \textbf{Model} & \textbf{LLaMA2-7B} & \multicolumn{2}{c|}{\textbf{Mixtral-8x7B}} & \multicolumn{2}{c}{\textbf{Qwen3-30B-A3B}} \\
    \midrule
    \textbf{Dataset} & \textbf{Dolly} & \textbf{Arxiv} & \textbf{GSM8K} & \textbf{Arxiv} & \textbf{GSM8K} \\
    \midrule
    Input Length & 22.25 & 8575.45 & 66.03 & 8050.69 & 61.51 \\
    Output Length & 246.87 & 227.08 & 126.79 & 208.57 & 121.03 \\
    \bottomrule
  \end{tabular}}
  \label{tab:lengths}
\end{table}

\begin{figure}[tb]
    \centering
    \includegraphics[width=0.48\textwidth]{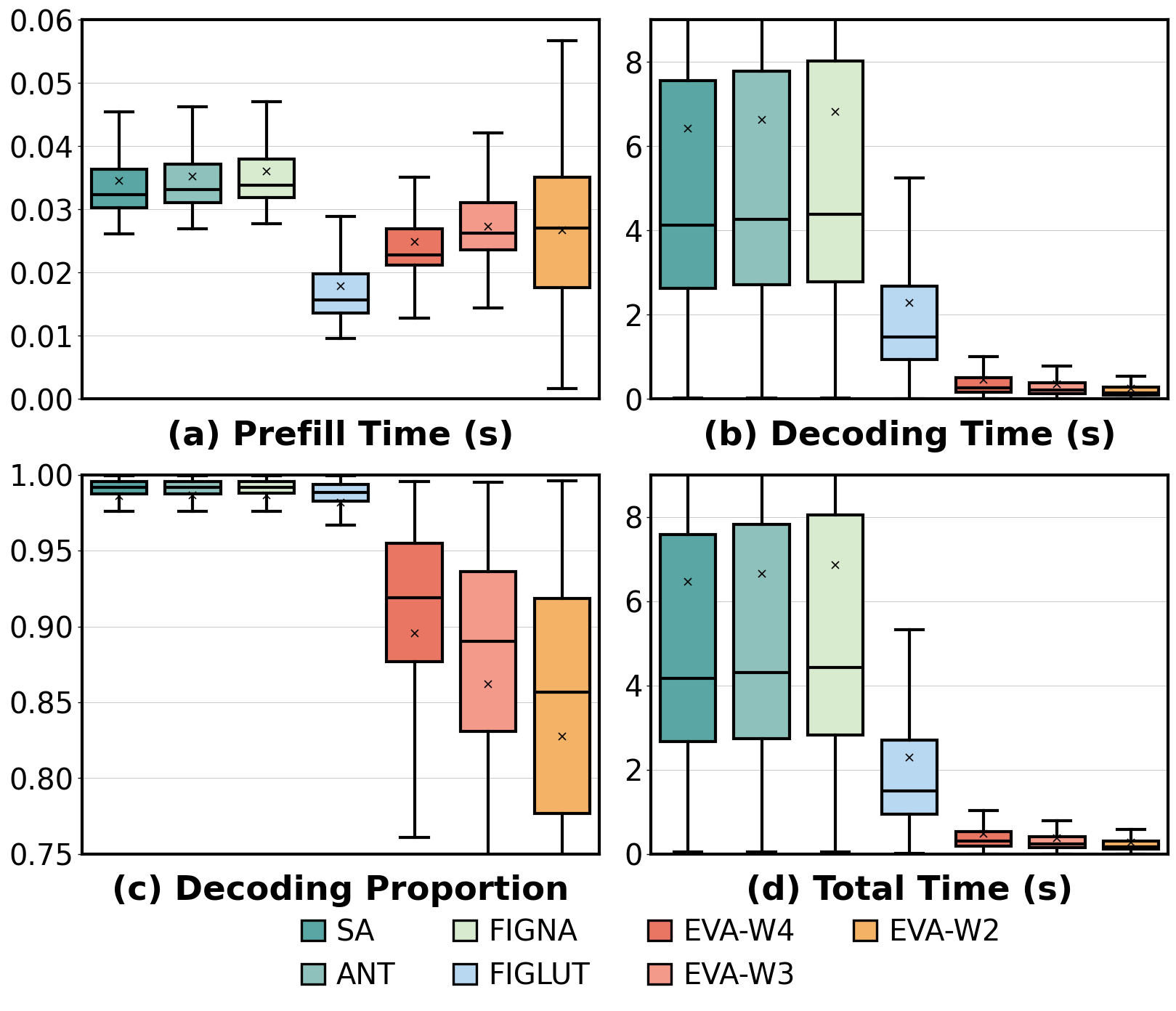}
    \vspace{-10pt}
    \caption{
    Results of running LlaMA-2-7B on the Dolly dataset using \eva{} and baseline accelerators. (a) Distribution of prefill time; (b) Distribution of decoding time; (c) Distribution of the ratio of decoding time to total time; (d) Distribution of the total end-to-end execution time.
    }
    \label{fig:dolly}
    \vspace{-5pt}
\end{figure}

\begin{figure}[tb]
    \centering
    \includegraphics[width=0.9\linewidth]{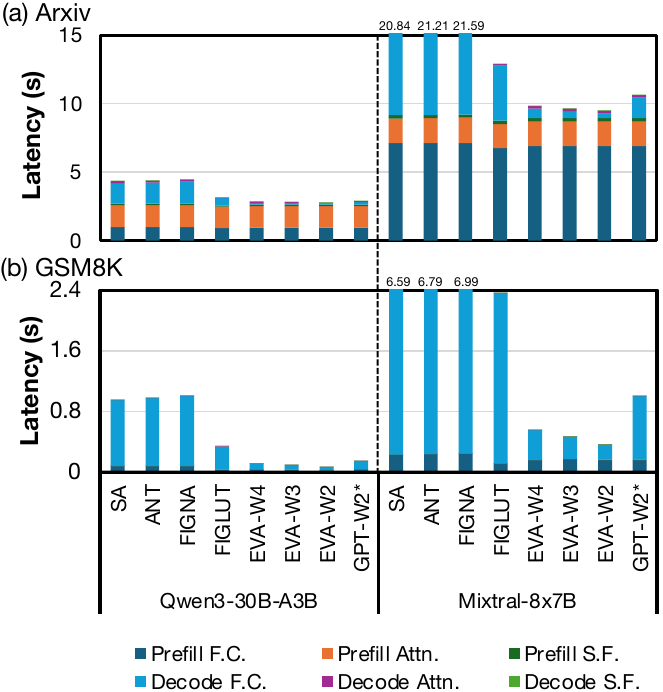} 
    \caption{\resp{CQ3}{End-to-end latency evaluation of MoE models on the (a) Arxiv Summarization and (b) GSM8K datasets. The results are categorized into prefill and decode phases for fully connected (F.C.), attention (Attn.), and special function (S.F.) layers. GPT-W2* represents the \eva{} architecture running the GPTVQ-4D algorithm.}}
    \label{fig:moe_e2e}
    \vspace{-3mm}
\end{figure}

\textbf{End-to-end performance on real-world datasets.} We evaluate the end-to-end performance of \eva{} and the baselines using LLaMA-2-7B model on the Dolly creative writing dataset~\cite{DatabricksBlog2023DollyV2}. Additionally, we evaluate the end-to-end performance of Mixtral-8x7B~\cite{jiang2024mixtral} and Qwen3-30B-A3B~\cite{yang2025qwen3} on the Arxiv Summarization~\cite{cohan2018discourse} and GSM8K~\cite{cobbe2021training} datasets. The input and output lengths for each dataset are detailed in \Tbl{tab:lengths}.

Following \Fig{fig:batch}, for \eva{}-A16W4/3/2, we use the $32\times32$ PE array to compute INT8 results when the input length exceeds 8, 16, or 32 tokens. The results in \Fig{fig:dolly} (a) show that all architectures have similar overall prefill latency. In \Fig{fig:dolly} (b), \eva{}-A16W2 again demonstrates clear decoding speed advantages, with 17.06$\times$ averaged speedup compared with all SOTA baselines. The Dolly dataset is decode-heavy. Therefore, decoding accounts for over 80\% of the total execution time across all architectures, as shown in \Fig{fig:dolly} (c). Consequently, \eva{}-A16W2 achieves an average 8.20$\times$ to 24.49$\times$ end-to-end speedup over the baselines, as illustrated in \Fig{fig:dolly} (d). These results highlight the importance of designing accelerators that efficiently support both prefill and decoding phases.

In MoE models (\Fig{fig:moe_e2e}), on the prefill-heavy Arxiv dataset, \eva{}-A16W2 achieves a 1.13$\times$--2.28$\times$ speedup over baselines. On decode-heavy GSM8K dataset, \eva{} significantly accelerates decoding, achieving a 5.01$\times$--18.92$\times$ speedup. 
Crucially, the breakdown proves attention is not the bottleneck. While attention consumes up to 59.53\% of runtime on the smaller Qwen model for Arxiv, this drops to 20.77\% for Mixtral, and becomes negligible (0.07\%--2.15\%) on GSM8K. Special functions also account for just 0.08\%--3.63\% of latency due to execution overlap, validating our optimization focus on F.C. layers. Finally, although the GPTVQ-4D configuration (\(2^n=256, N=256\)) inherently limits \eva{}'s performance (\Sec{sec:dse}), \eva{} running GPTVQ-4D still outperforms the SOTA FIGLUT baseline by 1.15$\times$ on Arxiv and 2.31$\times$ on GSM8K, confirming its effective support for diverse VQ algorithms.

\begin{table*}[t]
  \centering
  \small
  \renewcommand{\arraystretch}{0.6}
  \caption{\resp{CQ1}{Scaling Performance, Area Overhead, and Conflict Mitigation of \eva{} Configurations on LLaMA-2-7B.}}
  \resizebox{0.98\textwidth}{!}{
    \begin{tabular}{l | c c c | c c c}
      \toprule
      \textbf{Configuration\textbackslash Method} & \textbf{VQ w. Conflict} & \textbf{VQ-LLM} & \textbf{VQ w/o. Conflict} & \textbf{\eva{} EU-4 $\times$ 1} & \textbf{\eva{} EU-32 $\times$ 1} & \textbf{\eva{} EU-32 $\times$ 4} \\
      \midrule
      Codebook SRAM Size & \makecell{8 $\times$ 256 $\times$ \\ FP16 = \\ 4KB} & \makecell{8 $\times$ 256 $\times$ \\ FP16 $\times$ 2.5 = \\ 10KB} & \makecell{8 $\times$ 256 $\times$ \\ FP16 $\times$ 4 = \\ 16KB} & {\makecell{1 $\times$ 256 $\times$ \\ FP16 $\times$ 4 = \\ 2KB}} & \makecell{1 $\times$ 256 $\times$ \\ FP16 $\times$ 32 = \\ 16KB} & \makecell{1 $\times$ 256 $\times$ \\ FP16 $\times$ 32 $\times$ 4 = \\ 64KB} \\
      \midrule
      Codebook SRAM Bandwidth & \makecell{4 bank $\times$ 8 \\ $\times$ FP16 = \\ 64 B/Cycle} & \makecell{4 bank $\times$ 8 \\ $\times$ FP16 = \\ 64 B/Cycle} & \makecell{4 bank $\times$ 8 \\ $\times$ FP16 = \\ 64 B/Cycle} & {\makecell{4 bank $\times$ 1 \\ $\times$ FP16 = \\ 8 B/Cycle}} & \makecell{32 bank $\times$ 1 \\ $\times$ FP16 = \\ 64 B/Cycle} & \makecell{32 bank $\times$ 1 \\ $\times$ FP16 $\times$ 4 = \\ 256 B/Cycle} \\
      \midrule
      Systolic Array Size & 32$\times$8 (FP16) & 32$\times$8 (FP16) & 32$\times$8 (FP16) & 32$\times$8 (FP16) & {32$\times$8} (FP16) & 32$\times$8 (FP16) \\
      \midrule
      Epilogue Unit Size & - & - & - & \makecell{4-input \\ Adder} & \makecell{32-input \\ Adder} & \makecell{4 $\times$ 32-input \\ Adders} \\
      \midrule
      Normalized Array Area & 1.00$\times$ & 1.00$\times$ & 1.00$\times$ & {1.01$\times$} & 1.05$\times$ & {1.18$\times$} \\
      \midrule
      Normalized Speedup & 1.00$\times$ & 1.74$\times$ & 2.06$\times$ & \textbf{2.12$\times$} & \textbf{16.95$\times$} & \textbf{64.84$\times$} \\
      \midrule
      Note & Full Conflicts & 50\% Conflicts & No Conflicts & {No Conflicts} & No Conflicts & No Conflicts \\
      \bottomrule
    \end{tabular}
  }
  \label{tab:eva_scaling}
      \vspace{-10pt}
\end{table*}

\subsection{Discussion}
\label{sec:dis}
\textbf{\eva{}'s Output Codebook Advantages.}
We examine \eva{}'s architectural advantages within a $32\times8$ FP16 output-stationary systolic array configured for VQ ($d=8, n=8, C=1$), as illustrated in Table~\ref{tab:eva_scaling}. Standard dequantization retrieves vectors from codebook buffers divided into four banks to meet throughput requirements. However, simultaneous accesses to the same bank stall the array, introducing a $2.06\times$ latency overhead. To address this, we adapted the frequency-based replication of ``hot" weight indices from the GPU-optimized VQ-LLM framework~\cite{liu2025vq} into our simulator, which mitigates SRAM bank conflicts and yields a $1.74\times$ acceleration. In contrast, \eva{} fundamentally resolves this bottleneck by transforming weight-codebook lookups into output-codebook lookups. Because one Epilogue Unit (EU) accumulation implicitly replaces 8 MAC operations, this dimension collapse ($d=8 \rightarrow 1$) reduces both SRAM storage and bandwidth demands by $8\times$ under an identical 4-bank configuration. The final two columns of Table~\ref{tab:eva_scaling} demonstrate \eva{}'s scalability. 

\begin{figure}[t]
    \centering
    \includegraphics[width=\linewidth]{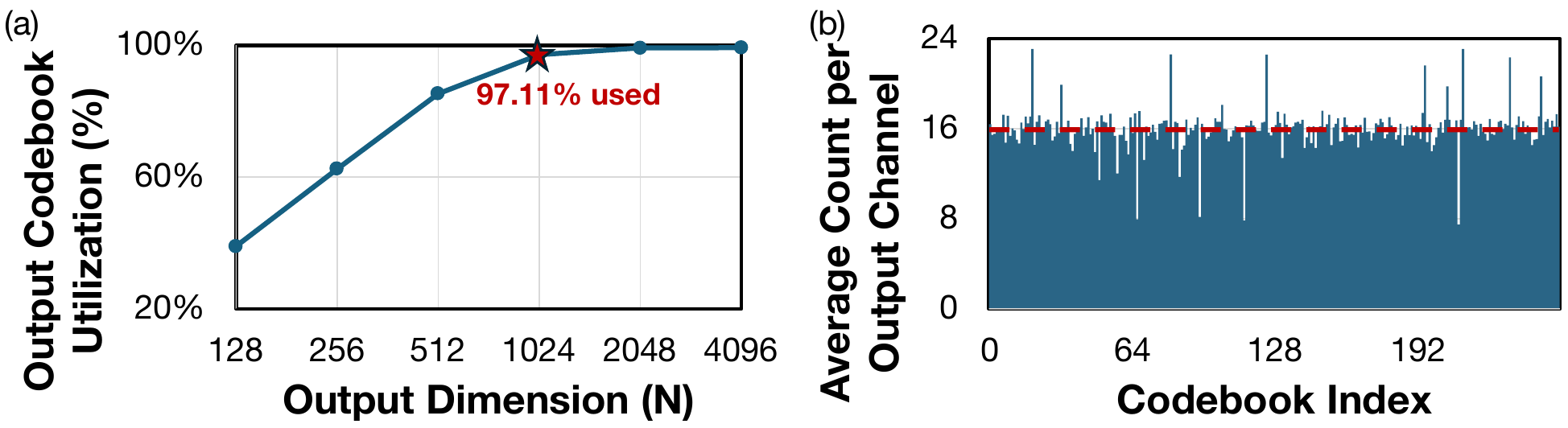} 
    \caption{\resp{RD-Q3}{Evaluation of spurious computations in the proposed method. (a) The effective computation rate (percentage of accessed codebook entries) scales with the number of output channels ($N$); (b) The access frequency of codebook indices averaged by output channel.}}
    \label{fig:spurious_comp}
    \vspace{-10pt}
\end{figure}

\textbf{Analysis of Spurious Computations.} In \eva{}, a multiplication is only ``spurious" if a centroid is computed but never referenced by any output channel \(N\). While redundancies can occur if the codebook size \(2^n\) exceeds \(N\), typical LLM layers feature \(N \gg 2^n\). Furthermore, the optimal VQ algorithm maximizes information entropy, naturally driving a uniform distribution of weight indices to prevent centroid collapse, as verified in \Fig{fig:spurious_comp}(b). Under this uniform distribution, the expected number of utilized centroids is \(\mathbb{E}[U] = 2^n \left[ 1 - \left(1 - \frac{1}{2^n}\right)^N \right]\). For \(2^n=256\) and \(N=1024\), the theoretical codebook utilization rate is \(98.2\%\), which closely matches our observed \(97.11\%\) in \Fig{fig:spurious_comp}(a). Therefore, centroid usage is inherently well-balanced, and the overhead from spurious multiplications is negligible.
\section{\resp{}{Related Work}}

\resp{RA-W2, RE-C3}{\textbf{Algorithm-level optimization for VQ-style LLM inference.}
Recent studies accelerate LLMs by leveraging VQ or similar methods.
VQ-LLM~\cite{liu2025vq} improves GPU execution of vector-quantized LLMs by identifying and optimizing frequently accessed codebook entries.
CodeGEMM~\cite{park2025codegemm} demonstrates the feasibility of VQ-based computation by reformulating quantized matrix operations for efficient GPU execution.
MADDNESS~\cite{blalock2021multiplying} proposes a hashing-based approximate matrix multiplication algorithm that replaces multiplications with learned lookup tables.
While these methods explore lookup-based computation in software, \eva{} co-designs the algorithm and architecture for LLM decoding.}

\resp{RC-W1, RE-C3}{\textbf{Lookup-based LLM accelerators.}
Several accelerators eliminate MAC operations through lookup-table computation.
FIGLUT~\cite{park2025figlut}, LUT Tensor Cores~\cite{zhiwen2024luttensorcore}, and Platinum~\cite{shan2026platinum} build lookup tables from activations and use weights as indices to skip multiplications in GEMM. Recent result-reuse accelerators, Prosperity~\cite{wei2025prosperity}, Transitive Array~\cite{guo2025transitive}, Phi~\cite{wei2025phi}, and Focus~\cite{wei2026focus}, enable lookup-based result reuse through activation sparsity.
While these designs eliminate arithmetic operations through lookup tables, they do not address the memory-system bottlenecks of VQ-based LLM inference; in contrast, \eva{} resolves memory bank conflicts to maximize compute utilization and is, to our knowledge, the first architecture-level accelerator for vector-quantized LLM inference.}
\section{Conclusion}
We presented \eva{}, a VQ-based architecture that restructures LLM decoding around a codebook-driven GEMM formulation. By directly multiplying inputs with the weight codebook and performing conflict-free lookups on an output codebook, \eva{} converts memory-bound GEMV into compute-efficient GEMM while eliminating codebook access conflicts. A mixed-precision GEMM unit and lightweight epilogue enable FP16 VQ decoding and INT8 prefill on a shared systolic array, achieving high utilization with modest hardware overhead. Evaluations on LLaMA-family models show that \eva{} sustains competitive accuracy at low bit-widths and delivers up to 11.17$\times$ speedup and 7.17$\times$ higher energy efficiency over the lookup-based baseline. Our work highlights the potential of codebook-aware, algorithm--hardware co-design for efficient LLM decoding.
\section{Acknowledgment}
This work was supported in part by NSF-2112562 and ARO W911NF-23-2-0224. The authors sincerely thank the anonymous reviewers for their constructive feedback and valuable suggestions that greatly improved the quality of this work. The authors also express their gratitude to Yuzhou Chen for his technical support and insightful discussions.


\bibliographystyle{IEEEtranS}
\bibliography{refs}
\appendix
\section{Artifact Appendix}

\subsection{Abstract}

Our artifact contains (1)~a hardware simulator that reproduces all hardware evaluation results (Figures~8--14 and Tables~III, VIII--IX) from the \eva{} paper, and (2)~an algorithm evaluation that reproduces the accuracy tables (Tables~V--VII) using pre-trained AQLM-quantized model checkpoints.

The hardware simulator models latency, energy, power, and area for \eva{} and baseline architectures (SA, ANT, FIGNA, FIGLUT) across dense and Mixture-of-Experts (MoE) LLMs. The algorithm evaluation runs perplexity and downstream benchmark evaluations on quantized LLMs hosted on Hugging Face. The artifact includes the full simulation pipeline, YAML-based study configurations, pre-processed trace files, evaluation scripts migrated from the AQLM repository~\cite{egiazarian2024extreme}, parallel shell scripts for batch reproduction, and Jupyter notebooks that render all paper-facing tables and figures.

\subsection{Artifact check-list (meta-information)}

{\small
\begin{itemize}
  \item {\bf Algorithm: } Hardware simulation of VQ-based \eva{} architecture for LLM inference; AQLM-based quantization evaluation.
  \item {\bf Program: } Python (NumPy, pandas, PyTorch, Transformers).
  \item {\bf Compilation: } Hardware simulator is interpreted in Python. Algorithm evaluation may JIT-compile AQLM CUDA/C++ extensions through \texttt{ninja}; GCC/G++ 11.x is recommended, and GCC/G++ 11.4.0 was tested.
  \item {\bf Transformations: } N/A.
  \item {\bf Binary: } N/A.
  \item {\bf Model: } LLaMA-2-7B, LLaMA-2-13B~\cite{touvron2023llama2}, Mixtral-8x7B~\cite{jiang2024mixtral}, Qwen3-30B-A3B~\cite{yang2025qwen3}. 
  \item {\bf Data set: } Dolly Creative Writing~\cite{DatabricksBlog2023DollyV2}, Arxiv~\cite{cohan2018discourse}, GSM8K~\cite{cobbe2021training}; WikiText-2 dataset~\cite{merity2016pointer}, PIQA~\cite{bisk2020piqa}, COPA~\cite{roemmele2011choice}, ARC-Easy, ARC-Challenge~\cite{clark2018think}, BoolQ~\cite{clark2019boolq}, and Winogrande~\cite{sakaguchi2021winogrande}.
  \item {\bf Run-time environment: } Linux (tested on Ubuntu 20.04+ and Ubuntu 22.04), Python 3.11, Conda, GCC/G++ 11.x for AQLM extension builds.
  \item {\bf Hardware: } Any x86-64 CPU with 16+ GB RAM (no GPU required for hardware simulation). Algorithm evaluation requires an NVIDIA GPU with $\geq$24\,GB VRAM (A100-80GB recommended).
  \item {\bf Run-time state: } Deterministic simulation (fixed seeds).
  \item {\bf Execution: } Python CLI scripts, parallel shell scripts, and Jupyter notebooks.
  \item {\bf Metrics: } Latency (cycles, seconds), energy (J), power (W), area (mm\textsuperscript{2}), throughput (GOPs), speedup; perplexity, accuracy (\%).
  \item {\bf Output: } CSV files containing per-study simulation results; JSON files with algorithm evaluation results; notebook-rendered tables and figures.
  \item {\bf Experiments: } 9 hardware simulation studies reproducing Figs.~8--14 and  Tables~III, VIII--IX; 10 algorithm evaluations reproducing Tables~V--VII.
  \item {\bf How much disk space required (approximately)?: } $\sim$10\,GB for hardware simulation; $\sim$100\,GB additional for algorithm evaluation (model checkpoints).
  \item {\bf How much time is needed to prepare workflow (approximately)?: } $\sim$5 minutes (Conda environment setup and package installation).
  \item {\bf How much time is needed to complete experiments (approximately)?: } $\sim$2 hours for hardware simulation (Steps~1--9); $\sim$6--8 hours additional for algorithm evaluation (Steps~10--12, on A100-80GB, depending on checkpoint download speed, GPU occupancy, and CUDA extension build/cache state).
  \item {\bf Publicly available?: } Yes. \url{https://github.com/dbw6/Eva.git}.  
  \item {\bf Code licenses (if publicly available)?: } MIT License.
  \item {\bf Data licenses (if publicly available)?: } Hugging Face dataset licenses (Apache 2.0 for Dolly; original licenses for Arxiv and GSM8K).
  \item {\bf Workflow automation framework used?: } N/A.
  \item {\bf Archived (provide DOI)?: } \url{https://doi.org/10.5281/zenodo.19433707}.
\end{itemize}
}

\subsection{Description}

\subsubsection{How to access}

The source code is publicly available at: \url{https://github.com/dbw6/Eva.git}.  
The artifact sources are also archived at Zenodo: \url{https://doi.org/10.5281/zenodo.19433707}.  
Pretrained weights for all evaluated models (LLaMA-2-7B, LLaMA-2-13B, Mixtral-8x7B, and Qwen3-30B-A3B) and datasets are available at: \url{https://huggingface.co/collections/dbw6/eva}.

\subsubsection{Hardware dependencies}
\textbf{Hardware Simulator:} Any x86-64 machine with at least 16\,GB of RAM and 10\,GB of free disk space. No GPU is required; all simulations run on the CPU. Internet access is required for the first run to download Hugging Face models and datasets.

\textbf{Algorithm Evaluation:} An NVIDIA GPU with at least 24\,GB VRAM (A100-80GB recommended), CUDA~12.x, and $\sim$100\,GB of additional disk space for downloading quantized model checkpoints.

\subsubsection{Software dependencies}
\begin{itemize}
    \item OS: Linux (tested on Ubuntu 20.04+)
    \item Python 3.11 and Conda (Miniconda or Anaconda)
    \item Compiler: GCC/G++ 11.x recommended; GCC/G++ 11.4.0 was used during artifact evaluation. AQLM may JIT-compile CUDA/C++ extensions through \texttt{ninja}, so the host compiler should be compatible with the installed CUDA toolkit.
    \item Core packages: \texttt{numpy}, \texttt{pandas}, \texttt{pyyaml}, \texttt{matplotlib}, \texttt{transformers}, \texttt{datasets}, \texttt{huggingface\_hub}
    \item Quantization support: \texttt{aqlm[gpu,cpu]>=1.1.6}
    \item Evaluation packages: \texttt{torch>=2.3.0}, \texttt{accelerate>=0.29.3}, \texttt{safetensors>=0.4.0}, \texttt{sentencepiece}, and \texttt{lm-evaluation-harness} 
\end{itemize}

\subsection{Installation}

Installation is performed via Conda and pip. The environment can be set up using the following commands:

\begin{verbatim}
conda create -n eva python=3.11 -y
conda activate eva
pip install -e .
pip install "aqlm[gpu,cpu]>=1.1.6"
pip install jupyter nbclient
\end{verbatim}

For the algorithm evaluation, additional PyTorch and Transformer dependencies must be installed as detailed in the \texttt{Eva/algorithm/README.md}.
If AQLM CUDA extension compilation fails, users should first verify that \texttt{gcc --version}, \texttt{g++ --version}, and \texttt{nvcc --version} are mutually compatible with the selected PyTorch/CUDA installation.

\subsection{Experiment workflow}

The evaluation is split into two independent workflows:

\textbf{1. Hardware Simulation (Steps 1--9):} Nine hardware simulation studies can be executed either one at a time via the CLI or with the provided parallel runner:
\begin{verbatim}
scripts/run_simulator_parallel.sh
\end{verbatim}

\textbf{2. Algorithm Evaluation (Steps 10--12):} Ten evaluations (4 perplexity + 6 downstream accuracy) can be run on a GPU to reproduce Tables~V--VII using the \texttt{eval\_ppl.py} and \texttt{lmeval.py} scripts. The provided multi-GPU runner distributes these jobs across a comma-separated GPU list and runs at most one evaluation per GPU:
\begin{verbatim}
scripts/run_algorithm_parallel.sh
\end{verbatim}

\subsection{Evaluation and expected results}

Each hardware study produces CSV files under \texttt{simulator/output/}, while algorithm evaluations produce JSON files under \texttt{algorithm/output/}. The provided Jupyter notebooks render these raw data files into the exact tables and figures presented in the paper. 

The total runtime for the hardware simulation is dominated by Step~9 (\texttt{e2e}), specifically the \texttt{fig13\_moe} scenario which simulates MoE models (Mixtral-8x7B and Qwen3-30B-A3B) across datasets with 100 samples each. Steps~1--8 complete in under 6 minutes combined. The full end-to-end hardware simulation takes approximately 2 hours. The algorithm evaluation takes approximately 6--8 hours on an A100-80GB GPU for a clean full run; multi-GPU execution with \texttt{scripts/run\_algorithm\_parallel.sh} can reduce wall-clock time when multiple suitable GPUs are available.

\subsection{Experiment customization}

Users can customize experiments by editing the YAML configuration files under \texttt{simulator/configs/studies/}. Each study YAML specifies models, methods, sequence lengths, batch sizes, and study-specific parameters. The CLI also accepts overrides via command-line arguments (e.g., \texttt{--models}, \texttt{--methods}, \texttt{--scenarios}).

\subsection{Methodology}

Submission, reviewing and badging methodology:
\begin{itemize}
  \item \url{https://www.acm.org/publications/policies/artifact-review-and-badging-current}
  \item \url{https://cTuning.org/ae}
\end{itemize}


\end{document}